\documentclass[onecolumn, draftcls]{IEEEtran}

\usepackage{amsmath}
\usepackage{bm}
\usepackage{epsfig}
\usepackage{amsfonts}
\usepackage{subfigure}
\usepackage{color}
\usepackage{cite}
\DeclareMathOperator{\sgn}{sgn}

\newcommand{\e}{\varepsilon}
\newcommand{\f}{\varphi}
\renewcommand{\Re}{{\rm Re}}
\renewcommand{\Im}{{\rm Im}}

\title{Unlimited-Power Reflectors, Absorbers, and Emitters with Conjugately Matched Layers}

\author{Constantinos A. Valagiannopoulos,\\
\textit{Department of Physics, School of Science and Technology, \\
Nazarbayev University, KZ-010000, Astana, Kazakhstan}\\
and \\
Sergei A. Tretyakov, \\
\textit{Department of Radio Science and Engineering, School of Electrical Engineering, \\
Aalto University, P.O. Box 13000, FI-00076 Aalto, Finland}}

\begin{document}
\maketitle

\begin{abstract}
In order to ensure the fastest wireless energy transfer from a source to the user one needs to maximize the channel capacity for power transport. In communications technologies, the concept of MIMO (multiple input, multiple output) exploits the idea of sending signals via many different rays which may reach the receiver. However, if we are concerned with the task of \emph{energy} transfer, still only one mode is exploited, even if multiple antennas are used to send power to the receiver. In the near-field scenario, this is the magnetic dipole mode of receiving coil antennas. In the far-field scenario, this is the propagating plane wave (transverse electromagnetic, TEM) mode. Recently, it was shown that using special artificial materials   it is possible to ensure that all electromagnetic modes of free space are conjugately matched to the modes of a material body and,  thus, all modes deliver power to the body in the most effective way. Such a fascinating feature is acquired because the conjugate matching does not concern only the propagating modes but, most importantly, is applied to all evanescent modes; in this way, all the possible ways of transferring the electromagnetic energy to the material body can be optimally exploited. However, coupling to higher-order (mostly evanescent) modes is weak and totally disappears in the limit of an infinite planar boundary. Here we show that properly perturbing the surface of the receiving or emitting  body with, for example, randomly distributed small particles we can open up channels for super-radiation into far zone. The currents induced in  the small particles act as secondary sources (radiation ``vessels'') which send the energy to travel far away from the surface and, reciprocally, receive power from far-located sources. We theoretically predict about 20-fold power transfer enhancement between the conjugately matched power-receiving body (as compared with the ideal black body) and far-zone sources. Reciprocally, the proposed structure radiates about 20 times more power into far zone as compared with the same source over a perfect reflector.

\end{abstract}

\maketitle

\section{Introduction}
The problem of optimizing and maximizing absorption and emission of electromagnetic energy is relevant for a broad variety or applications such as antennas, radar absorbers, thermal emitters and accumulators, photovoltaic devices and more. For macroscopic bodies (having sizes  large compared to the wavelength of electromagnetic radiation), it is usually assumed that the ultimate absorber and emitter is the ideal black body that completely absorbs all incident rays. Conceptually ideal black body, introduced by Kirchhoff \cite{GK}, is totally opaque and has zero reflection coefficient for any propagating plane wave (any incidence angle and any polarization), and in this sense it is the perfect absorber of electromagnetic energy. Following the Planck theory of thermal radiation \cite{Planck}, the ideal black body appears to be also the ultimate thermal emitter for radiating heat into free space. Practical realization of bodies whose properties mimic those of ideal black bodies is a scientific and technical challenge, see e.g. \cite{AbsorptionReview, OpticalBlackHole}.

However, it has been recently demonstrated that, in principle, it is possible to engineer bodies which can absorb power not only from incident propagating waves (incident rays in the Kirchhoff black-body concept), but also from external evanescent fields or high-order spherical harmonics of the incident-wave spectrum \cite{SphericalPaper, MAXABS}. Due to the resonant nature of surface modes of these superabsorbing and superemitting bodies, their absorption cross section grows without limit when the medium parameters approach the ideal values, and the thermal spectral emissivity at the resonance frequency becomes arbitrarily high compared to Planck's black body of the same size and the same temperature. The material structures proposed in \cite{SphericalPaper, MAXABS} realize the ideas of conjugate matching of all modes of free space to all modes of the absorbing/emitting body \cite{SoundBlackHole, Leman, unlimited_antenna}, and we call them \emph{conjugately matched} bodies or layers (CML).

The material realizations proposed in \cite{SphericalPaper, MAXABS} are based on the use of double-negative (DNG) isotropic or uniaxial media which obey the uniaxial perfectly matched layer (PML) conditions \cite{PML_Gedney}. High-order modes of conjugately matched bodies \cite{SphericalPaper, MAXABS} resonate with all modes of free space and most effectively exchange energy with them. However, to effectively absorb or emit all the modes, it is necessary that the modes of the body are sufficiently coupled with the corresponding modes of free space. In \cite{SphericalPaper}, it is shown that if a conjugately matched body fills a half space (a planar infinite interface with free space), its properties are the same as of the conventional ideal black body and no absorption or radiation enhancement over the ideal black-body limit can take place. To couple with higher-order free-space modes, we can make the surface not planar, and in \cite{SphericalPaper} it is shown that for bodies of finite sizes, on an example of a sphere, unlimited power exchange power is indeed possible. An alternative scenario was explored in paper \cite{MAXABS}, where the conjugately matched body filled a half space with an infinite planar interface, but the sources in free space were positioned close to the interface and created large evanescent fields which directly couple to the resonant surface modes of the conjugately matched layer. Also in this case it was seen that the absorption in the body was dramatically stronger than in an ideal black body at the same position.

In this paper, we show that it is possible to realize an infinite and planar surface which can absorb and emit more power than the ideal black body by perturbing the surface of a conjugately matched layer, introduced in \cite{MAXABS}. In this scenario, small subwavelength scatterers randomly distributed over the body surface, offer necessary coupling between high-order resonant surface modes and the far-zone fields, opening channels for extra absorption or emission of energy. In the limit of ideal material parameters, this planar interface not only absorbs or reflects all incident propagating waves, but does the same also for all evanescent harmonics. We show that a perturbed interface with a low-loss conjugately matched body acts as a ``super-reflector'' of fields created by a small antenna in its vicinity by launching the energy stored in the antenna near field into space. 

In particular, we introduce a random grid of electrically thin cylinders close to a resonant interface with a conjugately matched layer, where huge reactive energy is stored. Inevitably, currents induced in thin conductive cylinders radiate into far zone as linear antennas, and we say that these cylinders act as radiation ``vessels''. A random and sparse enough distribution of cylinders ensures that diffuse radiation survives in the far zone and is not coherently combined into a plane wave. We test the effects of this cluster of particles on the radiation from various conjugately matched layers and conclude that for a realizable passive structure one can achieve a stable 20-30-fold enhancement of the far-field power. 

Such super-reflectors are extremely strongly coupled to evanescent fields of external sources and can extract power from them in the most efficient way. In the antenna terminology, the effective area of the CML reflector is larger than the geometrical one, although the reflector size is very large compared with the wavelength. Basically, we aim at realization of a surface which (at its resonant frequency) would be ``more reflective than the ideal perfectly conducting mirror'', and this property would hold even in the limit of the infinite planar reflector. If the perturbing elements are lossy, instead of enhancing reflection we can enhance absorption in a planar absorbing layer or enhance thermal radiation from a planar hot surface into far zone beyond the Planck limit of the ideal black body. We expect that perturbing the surface can be a more effective mean to couple to far-zone field as compared to curved surfaces. In the study \cite{SphericalPaper} it was expectedly found that for large spherical bodies, when the curvature of the surface becomes small, one needs extreme (low loss) values of material parameters in order to realize enough effective coupling to high-order harmonics. Surface perturbation approach, introduced here, does not have this limitation.

\section{Conjugately Matched Layer (CML)}
We begin the study by a brief explanation of the concept of the conjugately matched layer, introduced in \cite{MAXABS}. It has been recently reported \cite{SphericalPaper, MAXABS} that there can exist material bodies which optimally absorb energy of electromagnetic fields, by achieving conjugate matching for every free-space mode. In the theoretical limit of negligible losses in the absorbing body, such an optimally designed finite-sized body can absorb the whole infinite energy of an incident propagating plane wave \cite{SphericalPaper}. In \cite{MAXABS}, a half-space and a cylinder filled with a uniaxial medium with special values for its constituent parameters have been suggested as possible realizations. The permittivities and permeabilities (transversal with subscript $t$ and normal to the material sample boundary with subscript $n$) satisfy the uniaxial perfectly matched layer (PML) conditions \cite{PML_Gedney,Ziolk,Text,synthetic,AriPaper} and simultaneously possess negative real parts as in a double negative (DNG) medium \cite{Veselago}, contrary to the conventional uniaxial PML choice. For planar interfaces and TM polarization (sole magnetic component parallel to the half-space boundary), the material parameters satisfy
\begin{equation}
\varepsilon_{rt}=\mu_{rt}=\frac{1}{\varepsilon_{rn}}=a-jb,
\label{PML_rule}
\end{equation}
where $a$ and $b$ are real parameters and $a<0$. The parameter $b>0$ corresponds to losses for propagating plane waves, and it is easy to show \cite{PML_Gedney,SergeiBook} that sufficiently thick slabs of such materials behave as perfect absorbers for arbitrary incident propagating plane waves. We assume harmonic time dependence $\exp(+j\omega t)$, where $\omega$ is the angular operating frequency. From duality, a similar expression for the parameters of uniaxial perfect absorbers holds for the fields of the TE polarization: $\varepsilon_{rt}=\mu_{rt}=\frac{1}{\mu_{rn}}=a-jb$. To ensure that the thought properties hold for both orthogonal polarizations, we can require that $\varepsilon_{rn}=\mu_{rn}$. For compactness, in the following we present only formulas for the TM polarization, without compromising the generality.

Any uniaxial medium characterized by the constituent parameters $(\e_{rt},\mu_{rt},\e_{rn})$, has the following TM wave impedance $Z$ (e.g., \cite{SergeiBook}):
\begin{equation}
Z=-j\frac{\eta_0}{k_0\varepsilon_{rt}}\sqrt{\frac{\varepsilon_{rt}}{\varepsilon_{rn}}k_t^2-\varepsilon_{rt}\mu_{rt}k_0^2},
\label{ZImpedance}
\end{equation}
which relates the tangential to the interface components of electric and magnetic fields of plane waves in the medium. The notations $\eta_0=\sqrt{\mu_0/\varepsilon_0}$ and $k_0=\omega\sqrt{\varepsilon_0\mu_0}=2\pi/\lambda_0$ correspond to the free-space wave impedance and wavenumber, respectively ($\varepsilon_0$ and $\mu_0$ are the permittivity and permeability of vacuum, while $\lambda_0$ is the operational wavelength in free space). The symbol $k_t$ is used for the transversal wavenumber of the incident plane wave. Vector $\textbf{k}_t$ is parallel to the boundary of the half space and normal to the sole component of magnetic field. The basic property of a material with the constituent parameters given by (\ref{PML_rule}) when $a<0$, is that its wave impedance $Z$ is the complex conjugate of the TM wave impedance of vacuum $Z^*=Z_0=-j\frac{\eta_0}{k_0}\sqrt{k_t^2-k_0^2}$~\cite{MAXABS}. 

Most importantly, this equality is valid for every real wavenumber $k_t\in\mathbb{R}$, either of a propagating wave in free space ($|k_t|<k_0$) or of an evanescent mode ($|k_t|>k_0$). Therefore, the use of such conjugate matched layers (CLM), as we call them, leads to maximal power transfer from arbitrary incident fields into the medium since they optimally use every possible way (mode) available from sources outside of the material sample. In particular, the TM impedance in vacuum $Z_0$ is real for propagating fields and accordingly $Z=Z_0$, which means that no power is reflected from the half-space surface, exactly as in the conventional uniaxial PML case \cite{PML_Gedney}. However, for evanescent fields we have a purely imaginary $Z_0$, which means that $Z=Z_0^*=-Z_0$. Thus, the maximal power transfer is achieved (in the limit of very small but non-zero losses in the medium) and simultaneously maximal reflections are taking place. Actually, in this ideal case of overall lossless conjugately matched  medium, fields tends to infinity at the material surface. Assuming infinitesimally small losses in the CML, infinite power can be delivered to the medium, provided that the incident evanescent field is created by an antenna fed by an ideal voltage or current source, capable of supplying infinite power. In other words, the CML structure is identical to ordinary PML for $|k_t|<k_0$ but operates totally differently for $|k_t|>k_0$ aiming not at zero reflection but at the maximal power transfer.

\begin{figure}[ht]
\centering
\subfigure[]{\includegraphics[scale =0.4]{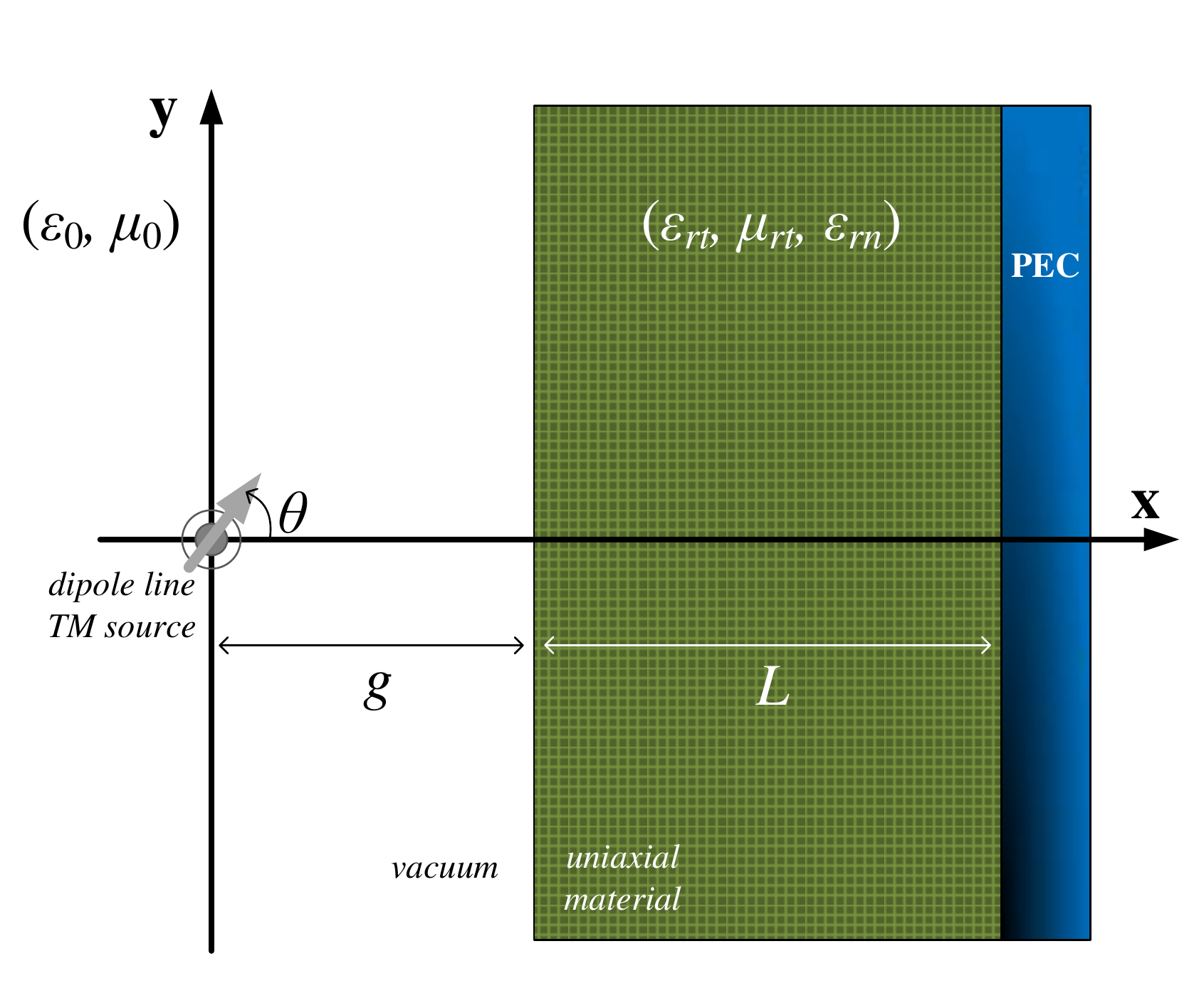}
\label{fig:Fig1a}}
\subfigure[]{\includegraphics[scale =0.4]{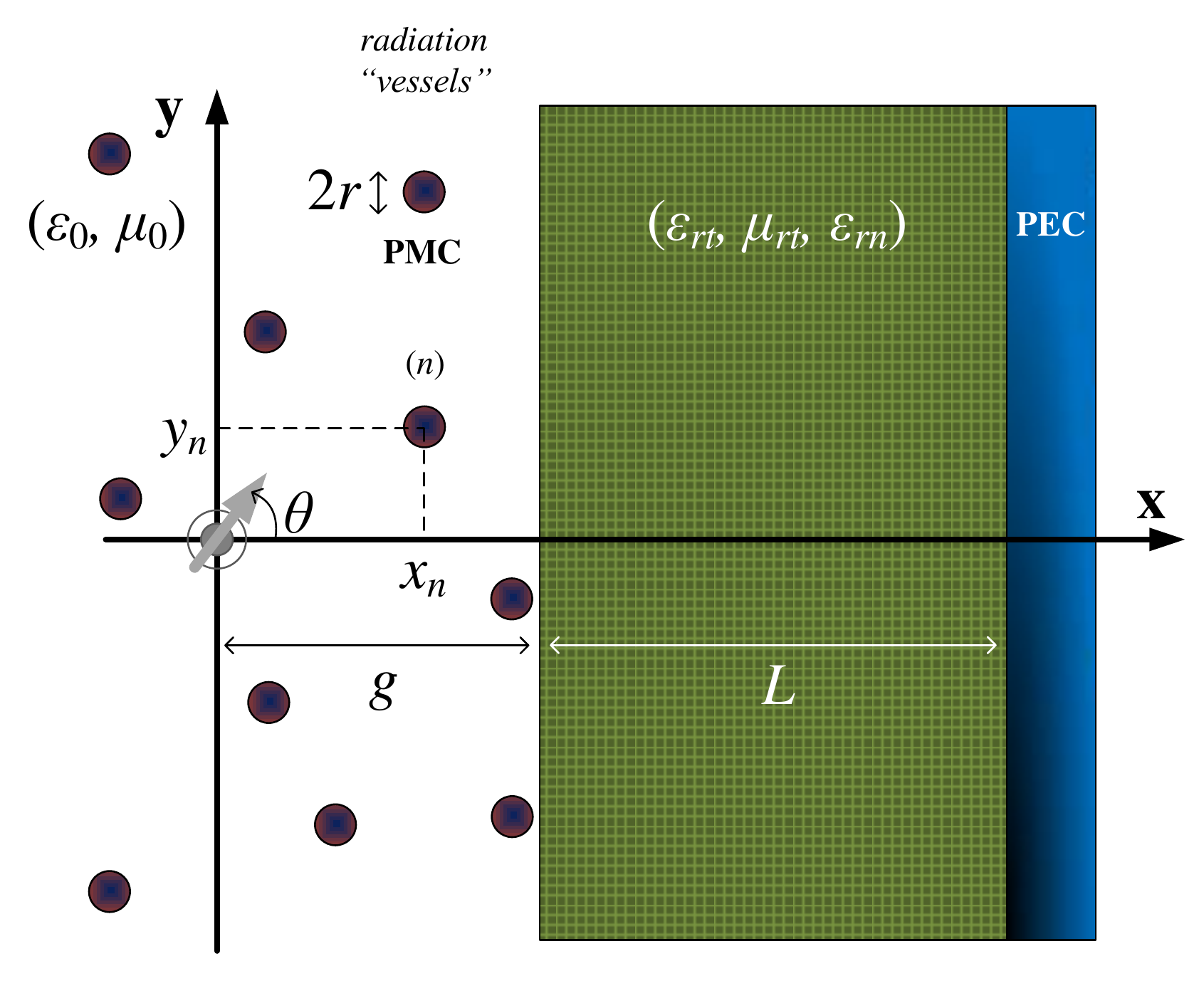}
\label{fig:Fig1b}}
\caption{(a) The test-bed configuration of a grounded electrically thick slab filled with a  uniaxial material $(\varepsilon_{rt},\mu_{rt},\varepsilon_{rn})$ with the thickness $L$,  excited by a TM electric dipole line inclined by the angle $\theta$ located at a small distance $g$ from the air-medium interface. (b) The same structure in the presence of a cluster of $N$ electrically thin (the radius $r$) circular perfect magnetic conductor (PMC) cylinders randomly distributed in the vicinity of the air-slab interface with arbitrary coordinates $(x_n,y_n)$ for $n=1,\cdots,N$.}
\label{fig:Figs1}
\end{figure}

By inspection of (\ref{PML_rule}) one can directly infer that if $b>0$, the transversal relative constituent parameters $\varepsilon_{rt},\mu_{rt}$, are lossy; however, the normal component of the permittivity $\varepsilon_{rn}$ is an active one. In order to identify the overall character of the uniaxial medium, we consider a perturbed version of the ideal material parameter values by using a small additional parameter $\delta$ controlling the imaginary part of the normal permittivity $\varepsilon_{rn}=\frac{1}{\varepsilon_{rt}}-j\delta=\frac{1}{a-jb}-j\delta$, which tends to the ideal CML medium parameters for $\delta\rightarrow 0$. To study the properties of such a quasi-CML medium, we use the test-bed configuration illustrated in Fig.~\ref{fig:Fig1a}. For simplicity of analytical considerations, we assume that there is no dependence on one of the tangential coordinates ($z$). A grounded slab of the thickness $L$, filled with a uniaxial material with the constituent parameters $(\varepsilon_{rt},\mu_{rt},\varepsilon_{rn})$, is excited by an infinite electric-dipole line located at the distance $g$ from the air-medium interface. The exciting dipole moments are orthogonal to the axis $z$ and inclined by the angle $\theta$ with respect to the axis $x$ (Fig.~\ref{fig:Fig1a}). We choose such a source since the spectral content for the evanescent modes is much more significant as compared to the common cylindrical wave (in the same way that has been used in \cite{DipoleLineSource}). In this configuration, the magnetic field has only one non-zero component (along $z$) and the fields are TM-polarized. 

In \cite{MAXABS}, an approximate analytical formula for the absorbed power (per unit length along the $\hat{\textbf{z}}$ axis) has been derived. It shows that the absorbed power is a sum of two terms. The first term corresponds to the power delivered by the propagating modes $P_{\rm prop}=\mu_0\omega^3q^2/16$, and it is independent from the angle $\theta$. Here $q$ is the electric dipole moment per unit length of the line (measured in $Coulomb$). The second term gives the power absorbed from the evanescent-modes fields and is given by \cite{MAXABS}
\begin{equation}
P_{\rm evan}\cong P_{\rm prop} \frac{8|a|}{k_0^2\pi} \int_{k_0}^{+\infty}\frac{k_t^2\left(k_t^2-k_0^2\sin^2\theta\right)}{\left(k_t^2-k_0^2\right)^{3/2}}e^{-2g\sqrt{k_t^2-k_0^2}}
\frac{\delta}{\left[1+\sgn(a)\right]^2+\delta^2\left[\frac{k_t^2 |\varepsilon_{rt}|}{2\left(k_t^2-k_0^2\right)}\right]^2}dk_t,
\label{EvanescentAbsorbedPower}
\end{equation}
which is an approximate expression valid under the assumption that $\delta\rightarrow 0$. It is remarkable that in the CML case ($a<0\Rightarrow \sgn(a)=-1$), the absorbed power from the evanescent-mode fields behaves as $P_{\rm evan}=O\left(1/\delta\right)$ for small $\delta$, namely, we obtain extremely high values of $P=P_{\rm prop}+P_{\rm evan}$, whose sign is the same as the sign of the perturbation variable $\delta$. In other words, the CML slab acts as an ultra-efficient passive absorber ($P\rightarrow +\infty$) of the incoming illumination for $\delta>0$ and as an infinite-power active emitter ($P\rightarrow -\infty$) for $\delta<0$. In the limit of $\delta\rightarrow 0^+$, both the field strength at the surface and the absorbed power diverge and tend to infinity. Therefore, it would be meaningful to inspect the field distributions leading to such unbounded field concentrations.

\section{Excitation of CML}
Let us examine the fields created by a small source in the vicinity of an infinite and homogeneous CML slab within the test-bed setup shown in Fig.~\ref{fig:Fig1a}. The corresponding boundary-value problem is scalar, and the magnetic field possesses a sole component parallel to $\hat{\textbf{z}}$ axis ($\textbf{H}=\hat{\textbf{z}}H(x,y)$). The used Cartesian coordinate system $(x,y,z)$ is also defined in Fig.~\ref{fig:Fig1a}, with the primary dipole line source positioned at $(x,y)=(0,0)$. The incident magnetic field from that electric-dipole line (existing in vacuum) can be expressed in the following integral form \cite{DipoleLineSource}:
\begin{equation}
H_{\rm inc}(x,y)=-\frac{\omega q}{4\pi}\int_{-\infty}^{+\infty}e^{-|x|\kappa_0(k_t)}
\left[\frac{k_t}{\kappa_0(k_t)}\cos\theta+j\sin\theta\sgn(x)\right]e^{-jk_ty}dk_t,
\label{IncidentField}
\end{equation}
where the normal to the interface component of the wavenumber $\kappa_0(k_t)=\sqrt{k_t^2-k_0^2}$ is evaluated with a positive real part and in case when the real part is zero, as a positive imaginary number. Analytical expression for the incident field involving Hankel function \cite{DipoleLineSource, Marcuvitz} is also available but not given here for brevity, since all the field quantities are expressed as spectral integrals. 

The formulated boundary-value problem can be solved analytically. As a result, we find that the secondary field developed due to the presence of the uniaxial slab and the PEC plane in free space ($x<g$) is given by: $H_{\rm sec}(x,y)=\int_{-\infty}^{+\infty}S_{\rm sec}(k_t)e^{\kappa_0(k_t)x-jk_ty}dk_t$, where
\begin{equation}
S_{\rm sec}(k_t)=\frac{\omega q}{4\pi}e^{-2g\kappa_0(k_t)}\left(\frac{k_t}{\kappa_0(k_t)}\cos\theta+j\sin\theta\right)
\frac{\kappa(k_t)-\varepsilon_{rt}\coth[\kappa(k_t)L]\kappa_0(k_t)}
{\kappa(k_t)+\varepsilon_{rt}\coth[\kappa(k_t)L]\kappa_0(k_t)}.
\label{SecondaryCoefficient}
\end{equation}
The value of 
\begin{equation}
\kappa(k_t)=\sqrt{k_t^2\frac{\varepsilon_{rt}}{\varepsilon_{rn}}-k_0^2\varepsilon_{rt}\mu_{rt}}
\label{beta}
\end{equation} 
is the normal component of the plane-wave propagation constant in the CML slab. The total field in vacuum equals to $H_{\rm back}(x,y)=H_{\rm inc}(x,y)+H_{\rm sec}(x,y)$.

\begin{figure}[ht]
\centering
\subfigure[]{\includegraphics[scale =0.5]{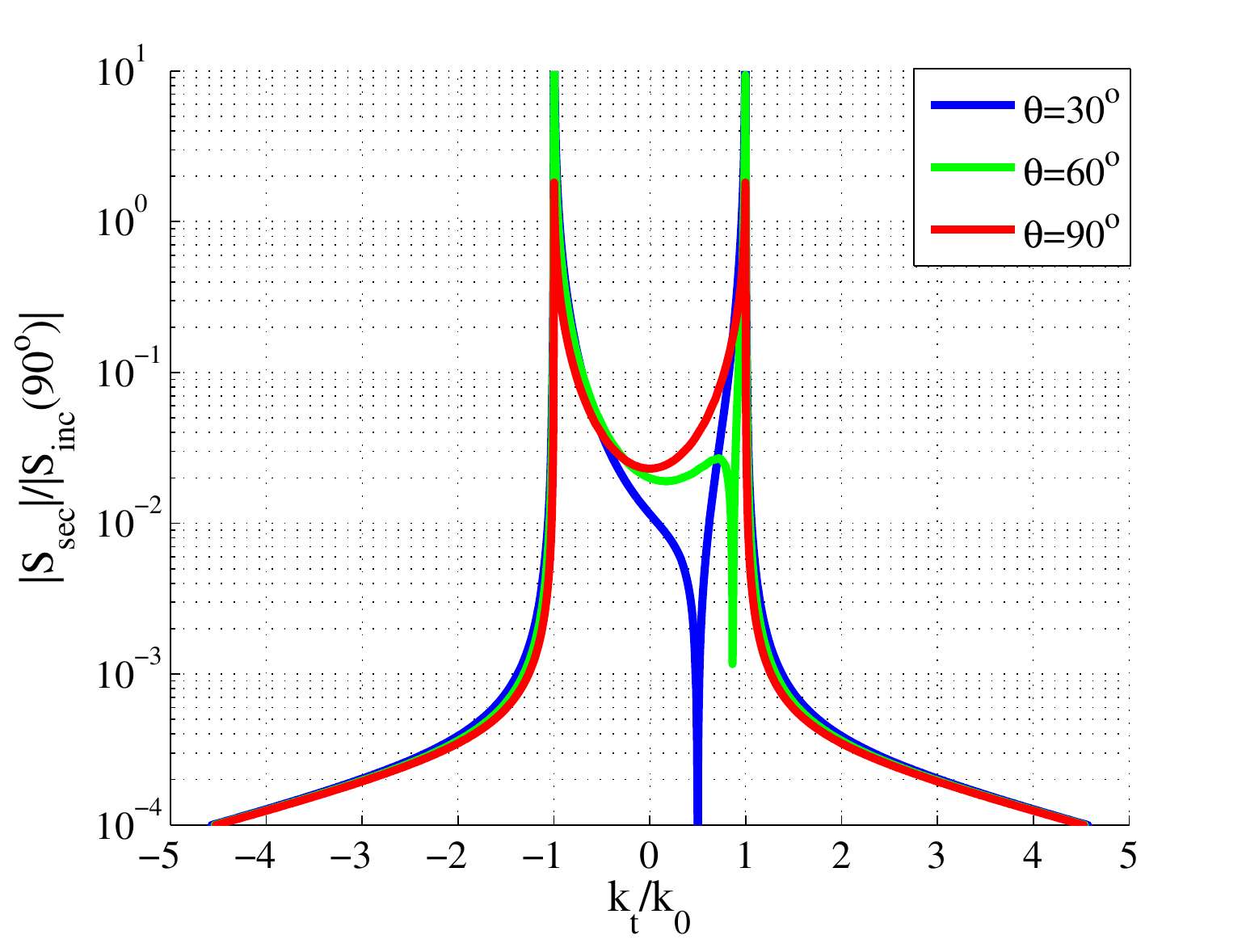}
\label{fig:Fig2a}}
\subfigure[]{\includegraphics[scale =0.5]{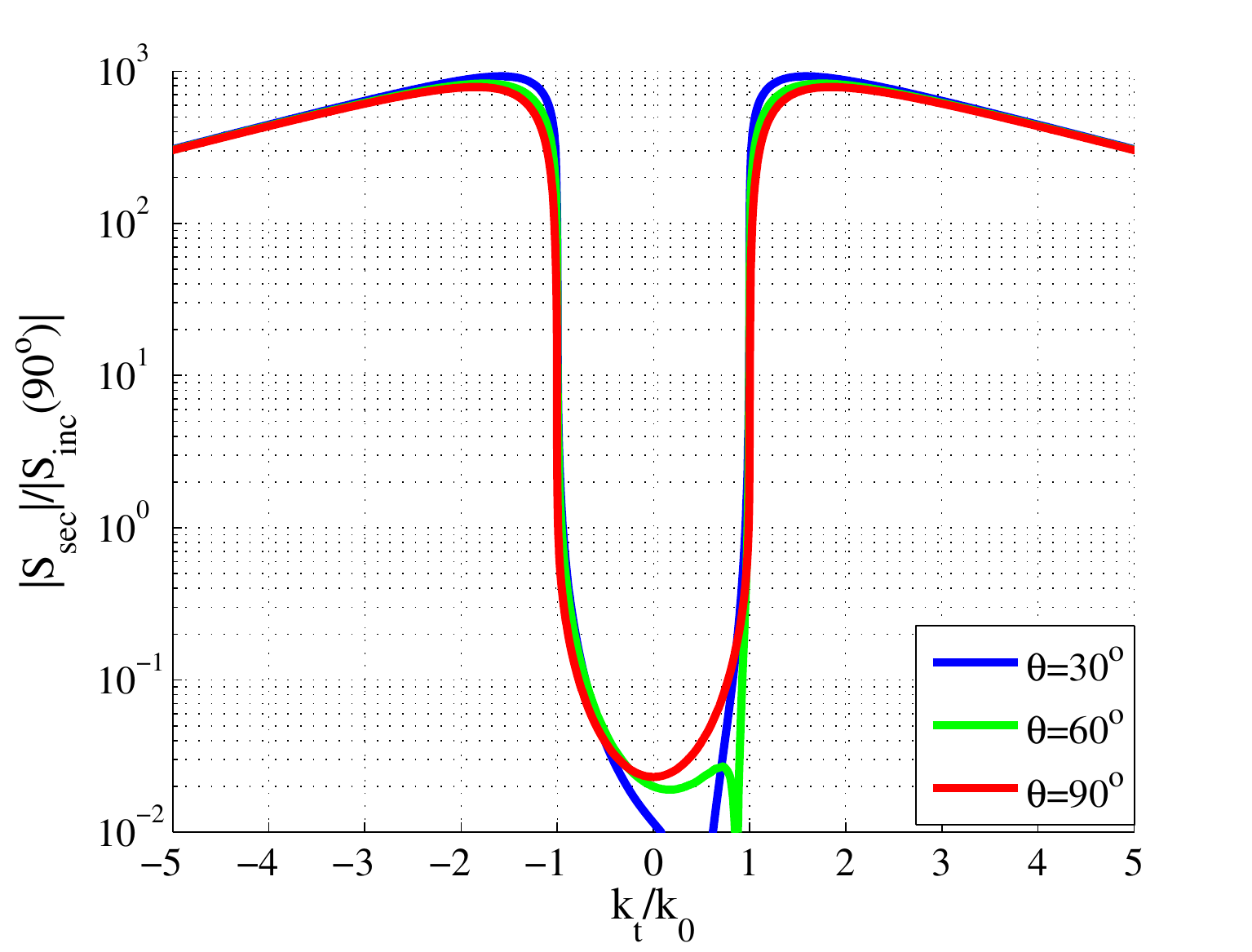}
\label{fig:Fig2b}}
\caption{The magnitude of the spatial spectral density function of the secondary field $S_{\rm sec}(k_t)$ at $x=0$ with respect to the normalized wavenumber $k_t/k_0$ for various inclination angles $\theta$ and: (a) $a=2$ (DPS-PML) and (b) $a=-2$ (CML). Common plot parameters: $b=0.1$, $\delta=0.001$, $g=0.03\lambda_0$, $L=3\lambda_0$. The represented quantity is normalized by  $S_{\rm inc}(90^{\circ})=j\omega q/(4\pi)$.}
\label{fig:Figs2}
\end{figure}

In Figs.~\ref{fig:Figs2} we present the magnitude of the integrand in the formula of the secondary magnetic field for $x=0$ (very close to the air-CML boundary located at $x=g$), equal to $|S_{\rm sec}(k_t)|$ as a function of the normalized transversal wavenumber $k_t/k_0$ for various inclination angles of the source $\theta$. The presented quantity is normalized by the (constant) magnitude of the integrand of the incident field (\ref{IncidentField}) for $\theta=90^{\circ}$: $S_{\rm inc}(90^{\circ})=\frac{j\omega q}{4\pi}$, which is independent from $k_t$ and gives us a metric of the incident power. Figure~\ref{fig:Fig2a} corresponds to a double-positive (DPS) conventional uniaxial PML \cite{PML_Gedney} and it is directly observed that $|S_{\rm sec}|$ vanishes exponentially for evanescent waves ($|k_t|>k_0$). On the contrary, for the CML case (Fig.~\ref{fig:Fig2b}) the integrand values have huge magnitudes for $|k_t|>k_0$ regardless of the angle $\theta$. These graphs verify the aforementioned theoretical expectation that unbounded absorbed power of (\ref{EvanescentAbsorbedPower}) in the CML case ($a<0$ and $|\delta|\rightarrow 0$) is due to the extremely large magnitudes of the evanescent fields developed in the vicinity of the interface, as demonstrated by Fig.~\ref{fig:Fig2b}. Note the different scale in Figs.~\ref{fig:Fig2a} and \ref{fig:Fig2b}: the values in the region $-1<k_t/k_0<1$, which correspond to the propagating-wave part of the spectrum, are the same in both figures.

\begin{figure}[ht]
\centering
\subfigure[]{\includegraphics[scale =0.5]{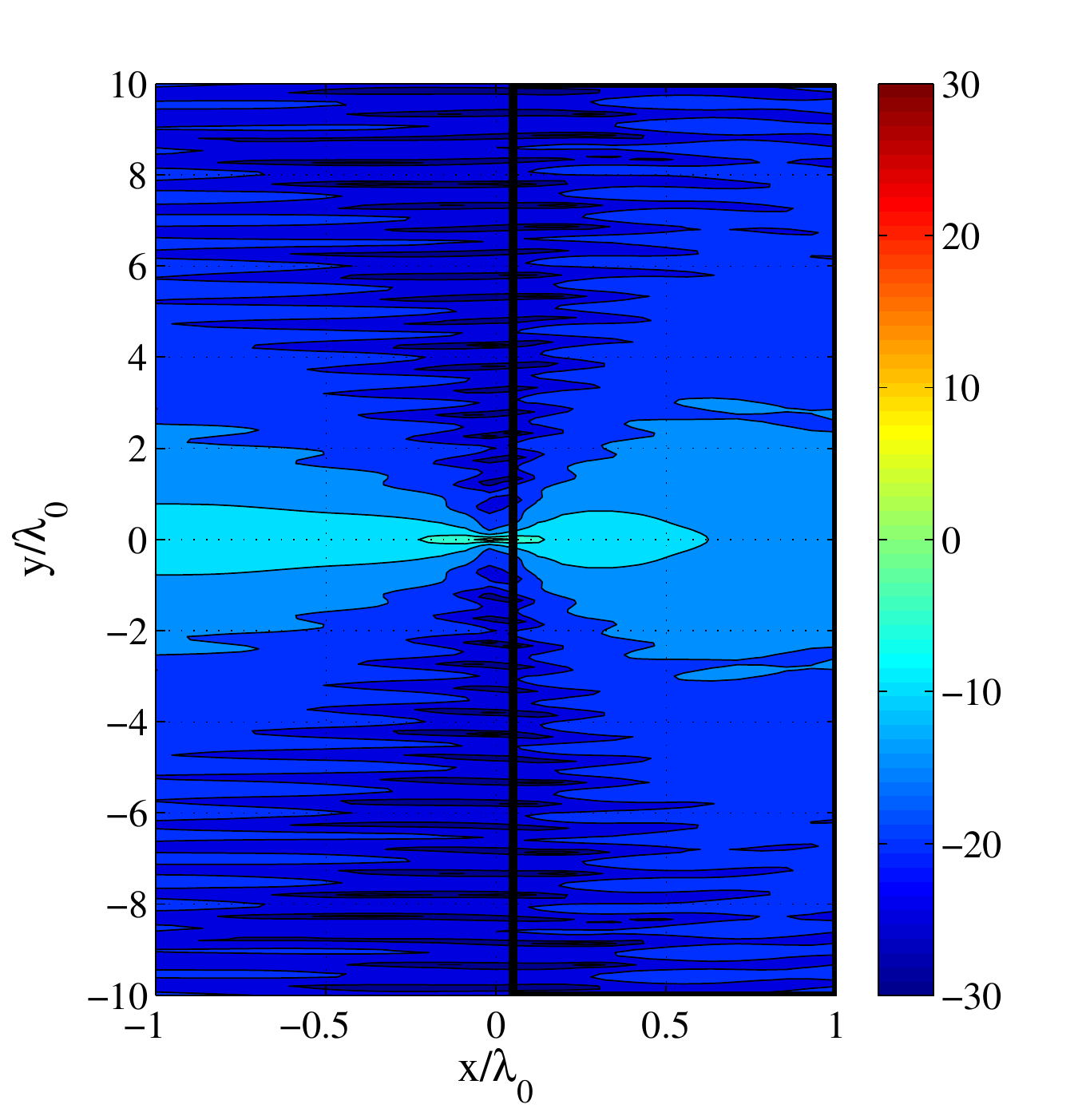}
\label{fig:Fig3a}}
\subfigure[]{\includegraphics[scale =0.5]{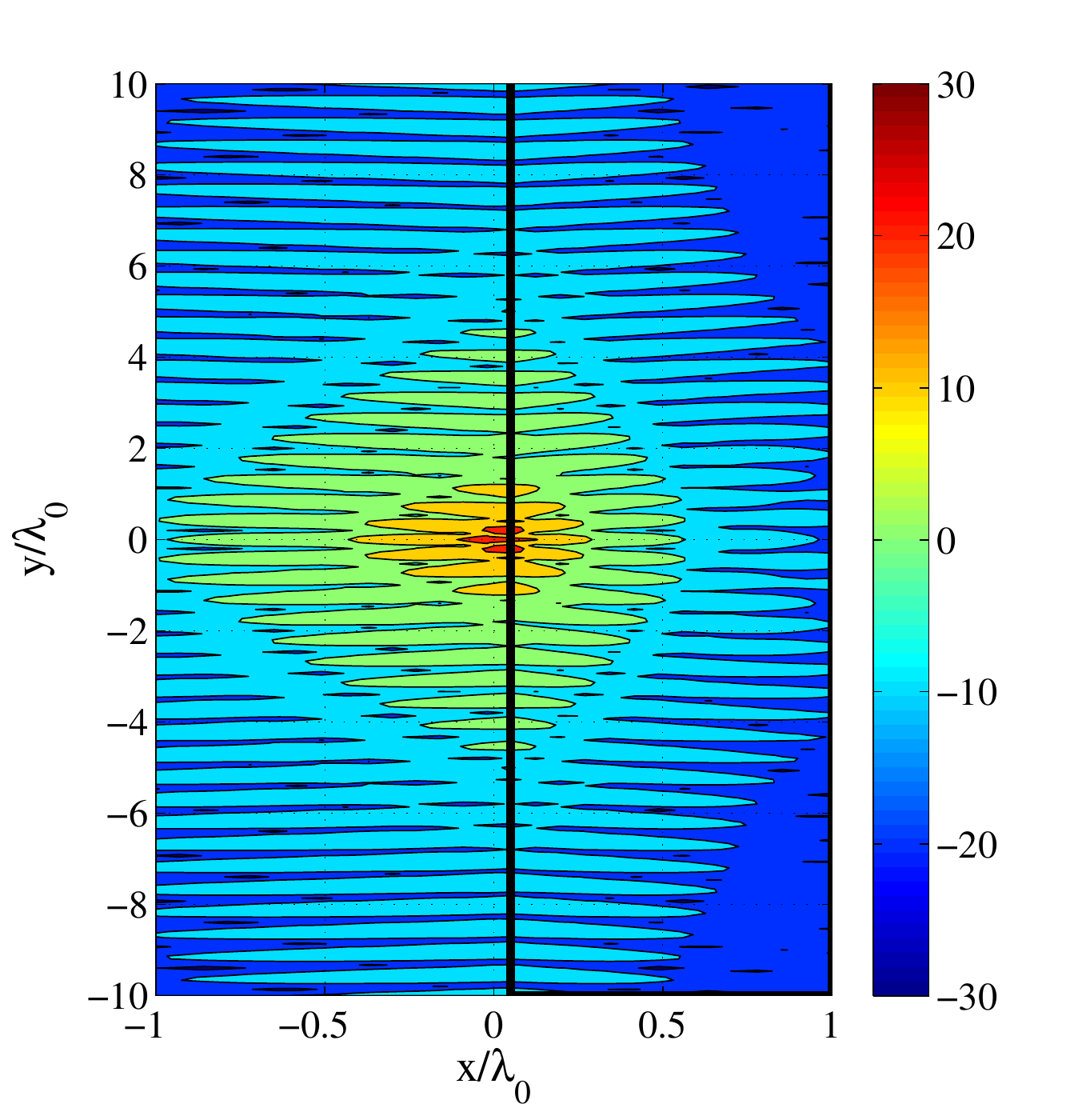}
\label{fig:Fig3b}}
\caption{The spatial distributions of the total magnetic field $H_{\rm back}(x,y)$ normalized by $H_{\rm inc}(g,0)$ expressed in $dB$ for: (a) $a=2$ (DPS-PML) and (b) $a=-2$ (CML). Common plot parameters: $b=0.1$, $\delta=0.001$, $g=0.03\lambda_0$, $L=3\lambda_0$, $\theta=90^{\circ}$.}
\label{fig:Figs3}
\end{figure}

Figures~\ref{fig:Figs3} show the spatial distribution of the total magnetic field $|H_{\rm back}(x,y)|$ for the two cases of Figs.~\ref{fig:Figs2} (with $\theta=90^{\circ}$). The represented quantity is normalized by the value of the incident field at $(x,y)=(g,0)$ and is expressed in $dB$. We again observe how more efficient is the CML medium (Fig.~\ref{fig:Fig2b}) in exciting fields along its boundary compared to the conventional PML case (Fig.~\ref{fig:Fig2a}). However, since the nature of these fields is evanescent, they are rapidly decaying with increasing the distance from the surface ($x\rightarrow -\infty$). It should be stressed that the concentration of the fields in the vicinity of the CML interface is always huge regardless of the sign of $\delta$, both for overall passive ($\delta>0$) or active ($\delta<0$) structures. 

In this paper, we propose to make use of this concentration of fields  along the boundary of the two regions ($x=g$) to create an ``antenna,'' which would ``launch'' the energy stored in this region into the far-zone region $x\rightarrow -\infty$. This is not an easy task, though. It is well-known that resonant surface modes at infinite and regular surfaces do not radiate energy into far zone. In other words, despite the huge difference of the two systems (PML versus CML slab) in the near field, the behavior of the field radiated in the far region is similar. Actually, with the use of the stationary phase method, one can directly evaluate the azimuthal profiles of the incident and the secondary fields in the far field as follows:
\begin{equation}
h_{\rm inc}(\varphi)\sim\frac{k_0j\omega q}{4}\sin(\varphi-\theta),~~~x\rightarrow-\infty,
\label{IncidentFarField}
\end{equation}
\begin{equation}
h_{\rm sec}(\varphi)\sim-\pi k_0 S_{\rm sec}(k_0\sin\varphi)\cos\varphi,~~~x\rightarrow-\infty,
\label{SecondaryFarField}
\end{equation}
for $90^{\circ}<\varphi<270^{\circ}$. We notice that the expression of the secondary component, which describes the effect of the grounded slab, is proportional to a specific value of the function $S_{\rm sec}(k_t)$: the one corresponding to $k_t=k_0\sin\varphi$. Since this value is always smaller in magnitude than $k_0$ ($\varphi\in\mathbb{R}$), namely corresponding to a propagating and not to an evanescent mode, it is clear that huge reactive fields of Fig.~\ref{fig:Fig2b} do not to contribute to far-zone radiation. We need something that can act as a radiation ``vessel'' to allow the field energy stored in resonant surface modes to propagate far away from the source.

\section{Radiation with ``Vessels''}
\subsection{Circuit Theory Approach}
In an attempt to find a way to exploit this huge field concentration and transform the sizeable magnitude of evanescent modes (developed close to $x=g$) into radiative fields, we consider the configuration of Fig.~\ref{fig:Fig1b}. Let us randomly distribute small cylindrical scatterers in the vicinity of the air-CML slab interface. It is expected that the large evanescent fields would excite currents along these wires, which will act as radiation vessels, and their own field would be expressed as cylindrical modes which are always partially propagating and survive in the far region. 

\begin{figure}[ht]
\centering
\includegraphics[scale =0.3]{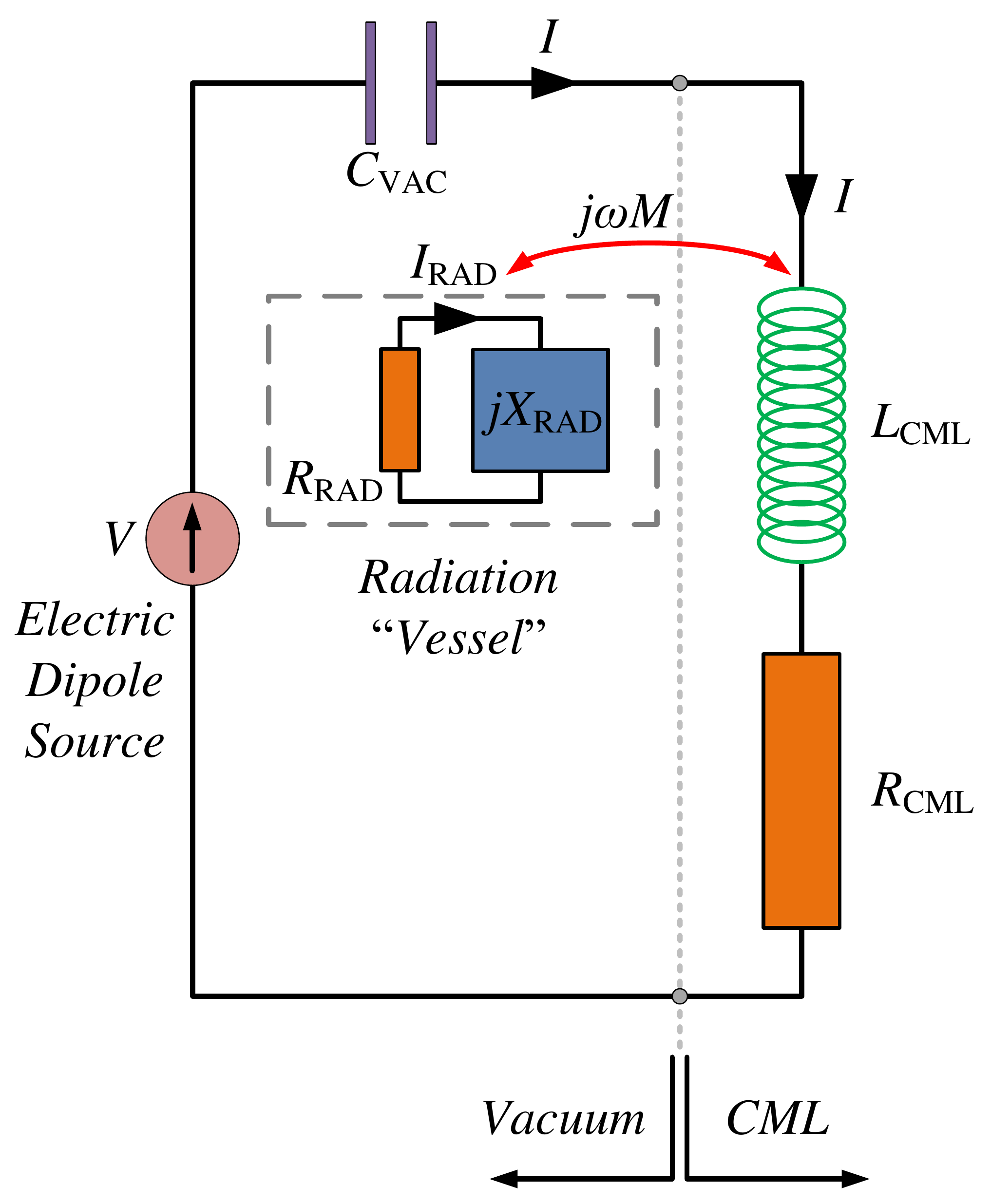}
\caption{Sketch of the equivalent circuit for excitation by a particular evanescent TM plane-wave component. The wave impedance of TM waves in vacuum $Z_0$ corresponds to the capacitance $C_{\rm VAC}$, while the impedance $Z$ of the CML contains a loss resistor $R_{\rm CML}$ and an inductive $L_{\rm CML}$ component. The cylindrical vessel in the near field of vacuum-CML interface is characterized by a radiation resistance $R_{\rm RAD}$ and a reactive impedance $X_{\rm RAD}$.}
\label{fig:Fig34}
\end{figure}

The idea of perturbing the surface with tiny scatterers can be understood from the equivalent circuit corresponding to the fields of a particular evanescent plane-wave component exciting the CML slab in presence of a small scatterer, shown in Fig.~\ref{fig:Fig34}. The ideal voltage source $V$ represents the primary radiator (an electric dipole source in this configuration) and capacitance $C_{\rm VAC}$ expresses the wave impedance of free space for a specific value of $k_t$, given by (\ref{ZImpedance}) for $\varepsilon_{rt}=\mu_{rt}=\varepsilon_{rn}=1$. The inductive complex impedance $(R_{\rm CML}+j\omega L_{\rm CML})$ is given by (\ref{ZImpedance}) with the parameters of the CML layer. Resistance $R_{\rm CML}$ models the dissipative losses in the CML slab and the current flown through the primary circuit is denoted by $I$. The small scatterer (radiation ``vessel'') in the vicinity of the interface is modeled by the radiation vessel circuit, of current $I_{\rm RAD}$, formed by a non-resonant reactive element $(jX_{\rm RAD})$ (capacitive or inductive) and a small resistor of the radiator $R_{\rm RAD}$. If the scatterer is lossless, $R_{\rm RAD}$ corresponds to the radiation resistance, and in case when it is absorptive, the resistance is the sum of the radiative and dissipative term. Near-field coupling between the scatterer and the resonant surface mode of the interface is modeled for simplicity by mutual inductance $j\omega M$. In general, the mutual impedance $j\omega M$ can be a complex number with any sign of the imaginary part; however, here we confine our analysis to a very closely positioned particle, in which case the mutual impedance is predominantly reactive ($M\in\mathbb{R}$) and inductive for the considered TM polarization.

The circuit in the absence of the radiation vessels $(M=0)$ has been analyzed in \cite{MAXABS}, and it is clear that the power delivered to the loss resistor $R_{\rm CML}$ tends to infinity when the series $LC$ circuit works at resonance and under the additional condition $R_{\rm CML}\rightarrow 0$. In other words, the absorbed power is limited only by the energy available from the primary source, while there is no radiation towards the far zone $(P_{\rm RAD}=0)$. In the presence of the vessels, however, the systems behaves dramatically different. Considering the circuit of Fig.~\ref{fig:Fig34}, we can easily find the current amplitude both in the directly fed branch ($I$) and in the circuit of the radiating vessel ($I_{\rm RAD}$):
\begin{equation}
I=\frac{V\left(R_{\rm RAD}+jX_{\rm RAD}\right)}
{\omega^2M^2+\left(R_{\rm RAD}+jX_{\rm RAD}\right)\left(R_{\rm CML}+j\omega L_{\rm CML}+\frac{1}{j\omega C_{\rm VAC}}\right)},
\label{CurrentI}
\end{equation}
\begin{equation}
I_{\rm RAD}=-\frac{j\omega M V}
{\omega^2M^2+\left(R_{\rm RAD}+jX_{\rm RAD}\right)\left(R_{\rm CML}+j\omega L_{\rm CML}+\frac{1}{j\omega C_{\rm VAC}}\right)}.
\label{CurrentIRAD}
\end{equation}

If we assume that there are no particles playing the role of radiation vessels ($M=0$), we notice that at the resonant frequency of the mode $j\omega L_{\rm CML}+\frac{1}{j\omega C_{\rm VAC}}=0$ and in the limit of negligible losses into the CML slab $R_{\rm CML}\rightarrow 0$, the current $I$ increases without bound. However, the radiated power $P_{\rm RAD}$ is zero because the resistance $R_{\rm CML}$ represents the dissipative process, not the radiative operation. In this way, we come again to the aforementioned conclusion that the power is accumulated in the near field and does not reach the far zone. On the contrary, when the pin comes close to the vacuum-CML interface, the radiated power equals to that delivered to $R_{\rm RAD}$ since it models the function of the vessel as antenna. Therefore,
\begin{equation}
P_{\rm RAD}=\frac{|V|^2 }{2}\frac{\omega^2 M^2 R_{\rm RAD}}{\left(\omega^2 M^2+R_{\rm RAD}R_{\rm CML}\right)^2+\left(X_{\rm RAD}R_{\rm CML}\right)^2},
\label{PRAD}
\end{equation}
under the assumption that the system works at CML resonant frequency, namely $\omega = 1/\sqrt{L_{\rm CML}C_{\rm VAC}}$. For resonant and low-loss CML ($R_{\rm CML}\rightarrow 0$), the expression for the radiated power simplifies to: $P_{\rm RAD}=\frac{R_{\rm RAD}|V|^2}{2\omega^2 M^2 }$. It is clear that in order to enhance radiation, we need to bring the CML to resonance and reduce its losses, while the vessels can be small and non-resonant. Coupling between the scatterers and the surface modes should be weak in the scenario. 

In the reciprocal situation of excitation by far-zone sources, we see that it is possible to enhance absorption beyond the ideal black-body full absorption of propagating plane waves by making the small scatterers lossy. In this case, assuming that the scatterers do not create a significant shadow for the propagating modes, the propagating plane waves deliver nearly all their power to the CML body, while the evanescent waves (high-order cylindrical harmonics) couple to the resonant surface modes via the small scatterers and deliver additional power into the loss resistors of the scatterers.

\subsection{Electromagnetic Theory Approach}
Having understood the basic operational principle from an  equivalent circuit, which is by default an approximation for every single mode $k_t$, we will next solve the problem rigorously for the entire spectrum of $k_t$. A spectrum integral  formulation is feasible if we assume random but specific positions of a finite number of scatterers $(x_n,y_n)$ on the $xy$ plane, where $n=1,\cdots, N$ (as shown in Fig.~\ref{fig:Fig1b}). For the sake of simplicity of test calculations, we assume that particles are circular cylinders of a small radius $r$ and of perfectly magnetically conducting (PMC) material. We chose perfect magnetic conductor pins as a simple model of lossless scatterers supporting magnetic currents, as appropriate for the considered TM polarization. Conceptual results will not change for any other small lossless scatterers at the same positions. Green's function of the considered configuration for both source $(\chi,\psi)$ and observation points $(x,y)$ in vacuum is comprised of two components. The singular component is just a cylindrical wave \cite{Marcuvitz}:
\begin{equation}
G_{\rm singular}(x,y,\chi,\psi)=-\frac{j}{4}H_0^{(2)}\left(k_0\sqrt{(x-\chi)^2+(y-\psi)^2}\right),
\label{SingularGreen}
\end{equation}
where $H_0^{(2)}$ is the Hankel function of zero order and second type. The smooth component of Green's function describes the effect of the grounded CML slab on the free-space field and is found as follows:
\begin{equation}
G_{\rm smooth}(x,y,\chi,\psi)=\int_{-\infty}^{+\infty}S_{\rm gre}(k_t)e^{\kappa_0(k_t)(x+\chi)}e^{-jk_t(y-\psi)}dk_t,
\label{SmoothGreen}
\end{equation}
where the spatial spectral density is given by \cite{MyOldWork}
\begin{equation}
S_{\rm gre}(k_t)=\frac{1}{4\pi}\frac{e^{-2\kappa_0(k_t)g}}{\kappa_0(k_t)}
\frac{\varepsilon_{rt}\coth[\kappa(k_t)L]\kappa_0(k_t)-\kappa(k_t)}{\varepsilon_{rt}\coth[\kappa(k_t)L]\kappa_0(k_t)+\kappa(k_t)}.
\label{SmoothGreenCoefficient}
\end{equation}

If we use the symbol $M_n$ ($n=1,\cdots, N$) for the magnetic currents (measured in $Volt/meter$) induced along the axes of the cylinders, the scattered magnetic field produced due to the presence of them is given as the following integral \cite{MFIE}:
\begin{equation}
H_{\rm scat}(x,y)=-\frac{jk_0}{\eta_0}\sum_{n=1}^N\int_{(C_n)}M_n(l)\left[G_{\rm singular}(x,y,\chi(l), \psi(l))+G_{\rm smooth}(x,y,\chi(l), \psi(l))\right]dl.
\label{ScatteredField}
\end{equation}
The notation $C_n$ is used for the contours of cylinder's surfaces. Since the cylinder radius is electrically small ($k_0r\ll 1$), the magnetic currents can be assumed to be uniformly distributed over the cylinder perimeter, and modeled by line currents along the cylinder axes, namely $M_n(l)\cong M_n$. In this way, the approximate boundary condition for zero total magnetic field at the centers of the cylinders $H_{\rm back}(x_m,y_m)+H_{\rm scat}(x_m,y_m)=0$ for $m=1,\cdots,N$ can be enforced to formulate the following $N\times N$ linear system of equations with respect to the unknown magnetic currents $M_n$:
\begin{equation}
\sum_{n=1}^N M_n \left[I_{mn}+2\pi r G_{\rm smooth}(x_m,y_m,x_n,y_n)\right]=\frac{\eta_0}{jk_0}H_{\rm back}(x_m,y_m).
\label{LinearSystem}
\end{equation}
The quantity $I_{mn}$ is the following approximately evaluated integral:
\begin{equation}
I_{mn}=\int_{(C_n)}G_{\rm singular}(x_m,y_m,\chi(l), \psi(l))dl=-\frac{j\pi r}{2}\left\{
\begin{array}{cc}
H_0^{(2)}(k_0r)       &, \; m=n\\
H_0^{(2)}(k_0d_{mn})  &, \; m\ne n
\end{array}
\right.,
\label{AuxiliaryIntegral}
\end{equation}
where: $d_{mn}=\sqrt{(x_m-x_n)^2+(y_m-y_n)^2}$ is the distance between the centers of the $n$-th and the $m$-th particle.

In this way, the induced magnetic currents can be found and the scattered field in the far region ($\varphi$-dependent profile) takes the form:
\begin{equation}
h_{\rm scat}(\varphi)\sim -\frac{j2\pi k_0r}{\eta_0}\sum_{n=1}^N 
M_n\left[e^{jk_0\rho_n\cos(\varphi-\varphi_n)}-\pi k_0 \cos\varphi S_{\rm gre}(k_0\sin\varphi)e^{-jk_0\rho_n\cos(\varphi+\varphi_n)}\right],~~~x\rightarrow-\infty,
\label{ScatteredFarField}
\end{equation}
where $\rho_n=\sqrt{x_n^2+y_n^2}$ and $\varphi_n=\arctan(x_n,y_n)$ are the polar coordinates of the cylindrical radiation vessels. Thus, we have obtained the analytical solution for the far field of the CML slab in the presence of numerous radiation vessels. In the following, we are going to use both approaches (circuit analysis and electromagnetic analysis) in order to study, interpret and quantify the radiation enhancement achieved when the pins are located in the vicinity of the vacuum-CML interface.

\section{Numerical Results}
In the following examples, we use a large number of vessels which are positioned neither very close to each other, to avoid building effective PMC walls which will block the incident illumination, nor too far since we want a strong background field at their positions. In particular, we locate $N=80$ random points $(x_n,y_n)$ for $n=1,\cdots, N$ belonging to a narrow vertical strip $\left\{-\lambda_0/20<x<\lambda_0/20~,-10\lambda_0<y<10\lambda_0\right\}$. The distance between every couple of centers of the cylinders $d_{mn}$ is kept larger than $\lambda_0/5$, so that the lattice is inhomogeneous at the wavelength scale and there is strong diffuse scattering into the far zone \cite{SergeiBook}. As referred above, we confine ourselves to uniaxial media (under TM illumination) with:
\begin{equation}
\e_{rt}=\mu_{rt}=a-jb~~~,~~~\e_{rn}=\frac{1}{\e_{rt}}-j\delta=\frac{1}{a-jb}-j\delta,
\label{CMLRule}
\end{equation}
and we are mainly interested in the CML cases with $a<0$. Obviously, the presented results are dependent on the random distribution of the radiation vessels; however, our studies of a number of particular realizations of the pins distribution show that the obtained conclusions are valid regardless of the spatial distribution of the PMC pins in the vicinity of the CML slab. 

\subsection{Radiation Enhancement}
A metric of how strong is the effect of the radiation vessels on the radiated far field strength should be definitely related with the energy of the azimuthal field profiles: $\left\{h_{\rm inc}(\varphi), h_{\rm sec}(\varphi), h_{\rm scat}(\varphi)\right\}$. In particular, we can define the radiation enhancement ratio $R$ as the ratio of the far-zone power radiated in the presence of the near-field scatterers and  the corresponding quantity in the absence of them:
\begin{equation}
R=\frac{\int_{\pi/2}^{3\pi/2}|h(\varphi)|^2d\varphi}
       {\int_{\pi/2}^{3\pi/2}|h_{\rm inc}(\varphi)+h_{\rm sec}(\varphi)|^2d\varphi}
\equiv\frac{\int_{\pi/2}^{3\pi/2}|h_{\rm inc}(\varphi)+h_{\rm sec}(\varphi)+h_{\rm scat}(\varphi)|^2d\varphi}
       {\int_{\pi/2}^{3\pi/2}|h_{\rm inc}(\varphi)+h_{\rm sec}(\varphi)|^2d\varphi}.
\label{RadiationEnhancementRatio}
\end{equation}
Here we evaluate and analyze the radiation enhancement factor $R$ when certain parameters of the analyzed configuration vary. We are seeking for combinations of the structure, the material parameters, and the excitation which lead to $R\gg 1$, namely a substantial improvement of the radiated power when one puts a random cluster of cylinders in the near region of the resonant surface.

\begin{figure}[ht]
\centering
\subfigure[]{\includegraphics[scale =0.43]{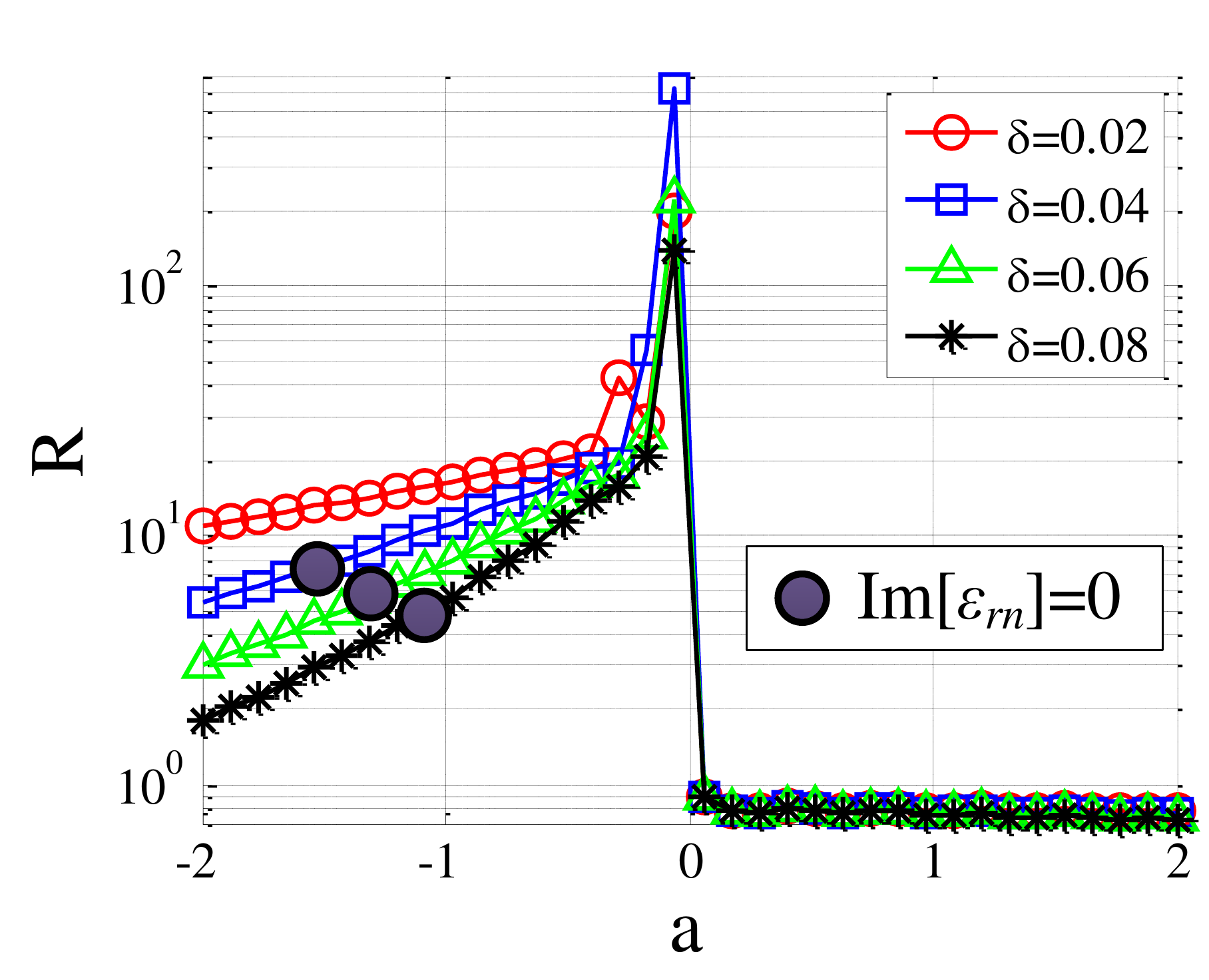}
\label{fig:Fig4a}}
\subfigure[]{\includegraphics[scale =0.43]{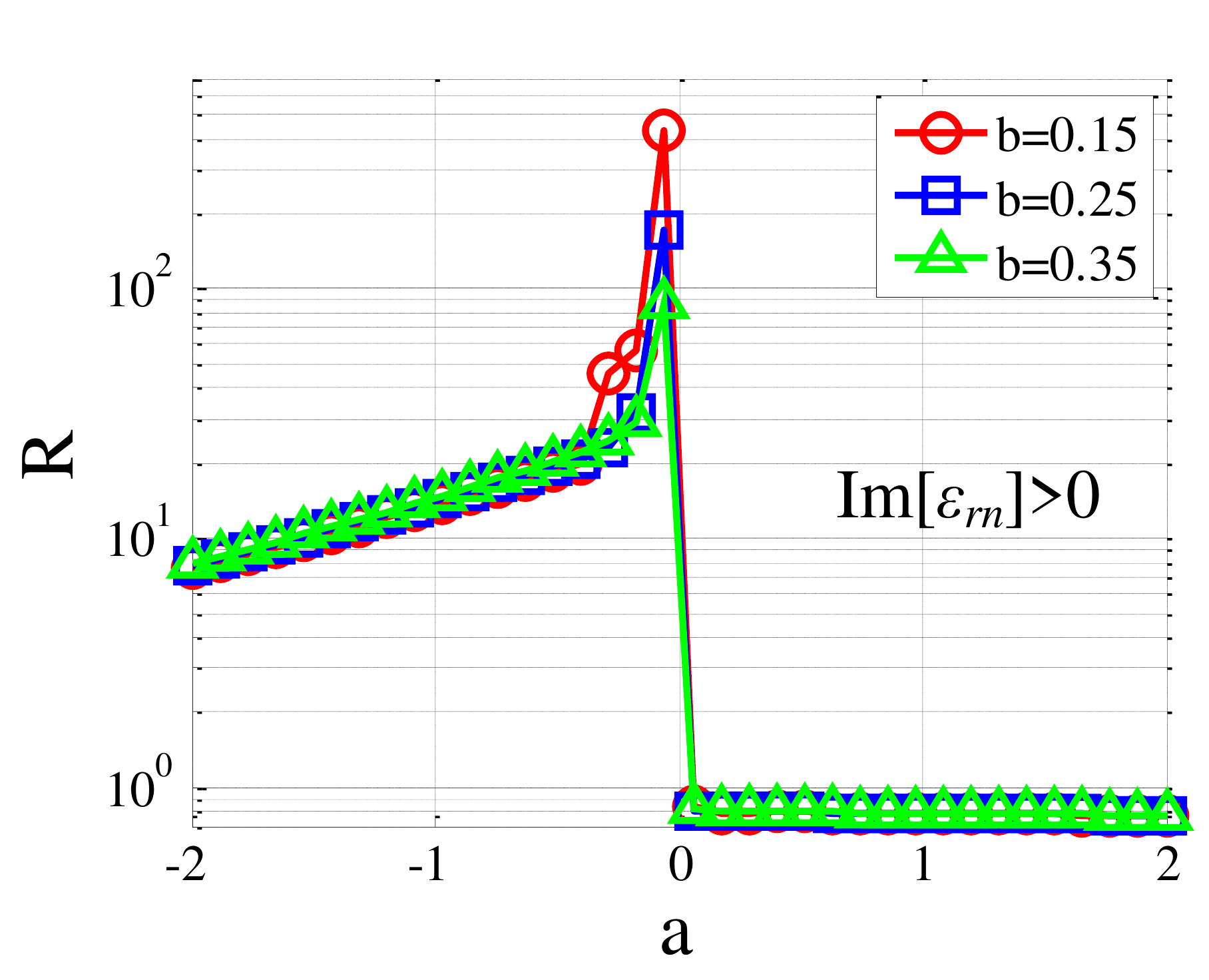}
\label{fig:Fig4b}}
\caption{The radiation enhancement ratio $R$ as a function of the real part of the transversal components $a=\Re[\e_{rt}]=\Re[\mu_{rt}]$ for: (a) several perturbation parameters $\delta=\Im[1/\e_{rt}-\e_{rn}]$ of the normal component of the material parameters (with $b=0.1$) and (b) several values of the imaginary part $b=-\Im[\e_{rt}]=-\Im[\mu_{rt}]$ (with $\delta=0.03$). Plot parameters: $r=\lambda_0/200$, $N=80$, $g=\lambda_0/20$, $L=3\lambda_0$,  $\theta=90^{\circ}$.}
\label{fig:Figs4}
\end{figure}

In Fig.~\ref{fig:Fig4a} we show the ratio $R$ as a function of the real part $a$ of the relative transversal permittivities/permeabilities ($a=\Re[\e_{rt}]=\Re[\mu_{rt}]$) for various perturbation parameters $\delta$. One directly observes a huge change in the magnitude of $R$ taking place when the material parameters transit from the double-negative (CML) slabs ($a<0$) to double-positive, conventional PML slabs ($a>0$). This feature is explained by the resonant nature of the CML with $a<0$. That is why we are focusing on the case of CML ($a<0$) rather than the conventional uniaxial PML ($a>0$). With the purple dots we show (in the DNG cases $a<0$) the points on each curve for which the normal permittivity becomes lossless (it is lossy on the left side of the dots and active on the right side). In other words, the dots indicate the equality: $\delta=\frac{b}{a^2+b^2}\Rightarrow a=-\sqrt{\frac{b}{\delta}-b^2}$, which is valid (within the considered ranges of $a$) only for the three of the four curves of Fig.~ \ref{fig:Fig4a} (and for none of Fig. \ref{fig:Fig4b}). It is clear that when one moves along that ``ultimate passivity boundary'' (where none of the permittivity/permeability components is active)  defined  by the aforementioned successive purple dots, takes a decreasing $R$ both for increasing $\delta$ and for increasing $a<0$ (when $|a<0|$ becomes smaller). We can also conclude that even when the medium is lossy for any direction of the fields, the radiation enhancement is significant. 

Most importantly, these results prove that the radiation enhancement due to strong coupling of resonant surface modes to the far-field modes is orders of magnitude stronger than possible reduction of radiation of propagating modes into far zone. Recall that in the absence of the scatterers and $\delta\rightarrow 0$ the CML slab is perfectly matched to free space. For small values of $b$, all propagating modes are fully reflected, and we clearly see that adding pins makes the reflected fields more than two orders of magnitude  stronger than reflected from a  conventional perfect reflector.   Furthermore, $R$ is larger when $\delta$ is closer to zero which is anticipated by the limiting expression of (\ref{EvanescentAbsorbedPower}) in the DNG case. The best results are recorded when $a$ is negative but close to zero (much larger than $-1$, namely for $-1\ll a<0$) where the radiation enhancement is giant and practically independent from $\delta$. In Fig.~\ref{fig:Fig4b}, we represent $R$ as a function of $a$ for several loss parameters $b$. Again, we note that the cluster works only in the CML case, where it does a very good job ($R>50$ on average). Finally, a smaller imaginary part $b$ (with fixed $\delta>0$) favors the increase in the radiated power achieved with the cylindrical vessels.

\begin{figure}[ht]
\centering
\includegraphics[scale =0.43]{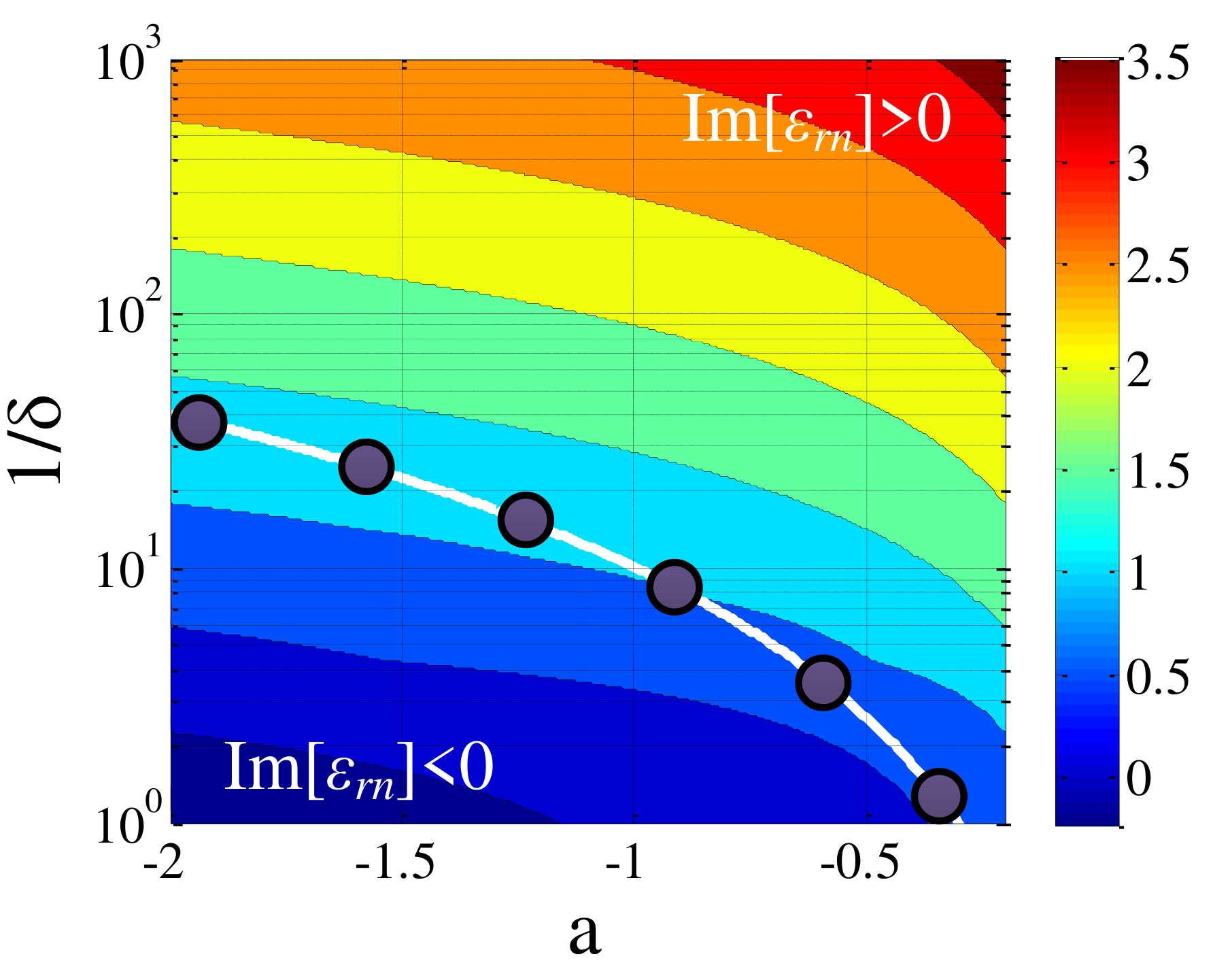}
\caption{The decimal logarithm of the quality factor of the equivalent circuit $\log Q$ with respect to the real part of the transversal components $a=\Re[\e_{rt}]=\Re[\mu_{rt}]$ and the inverse perturbation parameter $1/\delta=1/\Im[1/\e_{rt}-\e_{rn}]$. Plot parameters: $k_t=1.5k_0$, $b=0.1$.} 
\label{fig:Fig45}
\end{figure}

The singular behavior of the radiated power in the limit $a\rightarrow 0^-$ can be explained by considering the quality factor of the resonating surface modes. To do that, we calculate the equivalent inductance $L_{\rm CML}$ and resistance $R_{\rm CML}$ considering the wave impedance $Z$ of the CML medium (\ref{ZImpedance}); similarly one can find expressions for the capacitive effect of free space $C_{\rm VAC}$ by evaluating $Z_0$. If one assumes that $\delta>0$ (to ensure overall passivity), evanescent modes $|k_t|>k_0$ (for which the interesting phenomena happen) and $a<0$ (to have resonance), the quality factor of the  equivalent $RLC$ series circuit takes the form:
\begin{eqnarray}
Q=\frac{1}{R_{\rm CML}}\sqrt{\frac{L_{\rm CML}}{C_{\rm VAC}}}\Rightarrow
Q\cong\frac{\sqrt{2(k_t^2-k_0^2)}}{k_t^2}\frac{\sqrt{2(k_t^2-k_0^2)+b\delta k_t^2}}{(-a)\delta}~,~~~\delta\rightarrow 0^+.
\label{QFactor}
\end{eqnarray}
It is easy to see that the loss parameter $R_{\rm CML}$ is proportional to $(-a)\delta$ in this case. Thus, for a fixed level of losses in the CML slab (measured by $\delta$), the quality factor behaves as $1/a$ for $a\rightarrow 0^-$. Figure~\ref{fig:Fig45} shows the values of the quality factor on the plane $(a,1/\delta)$ in the region $-2<a<0.2$, $1<1/\delta<100$. It is clear that $Q$ obtains huge magnitudes when $|a|, \delta$ are very small for the CML scenario, namely under the assumption of $a<0$. The ultimate passivity boundary, along which we have a nonactive $\e_{rn}$ ($\Im[\e_{rn}]=0$), is indicated by a white line with purple dots. It divides the map $(a,1/\delta)$ into two regions: one upper right which corresponds to active normal permittivity ($\Im[\e_{rn}]>0$) and one lower left which concerns a passive normal permittivity ($\Im[\e_{rn}]<0$).

\begin{figure}[ht]
\centering
\subfigure[]{\includegraphics[scale =0.43]{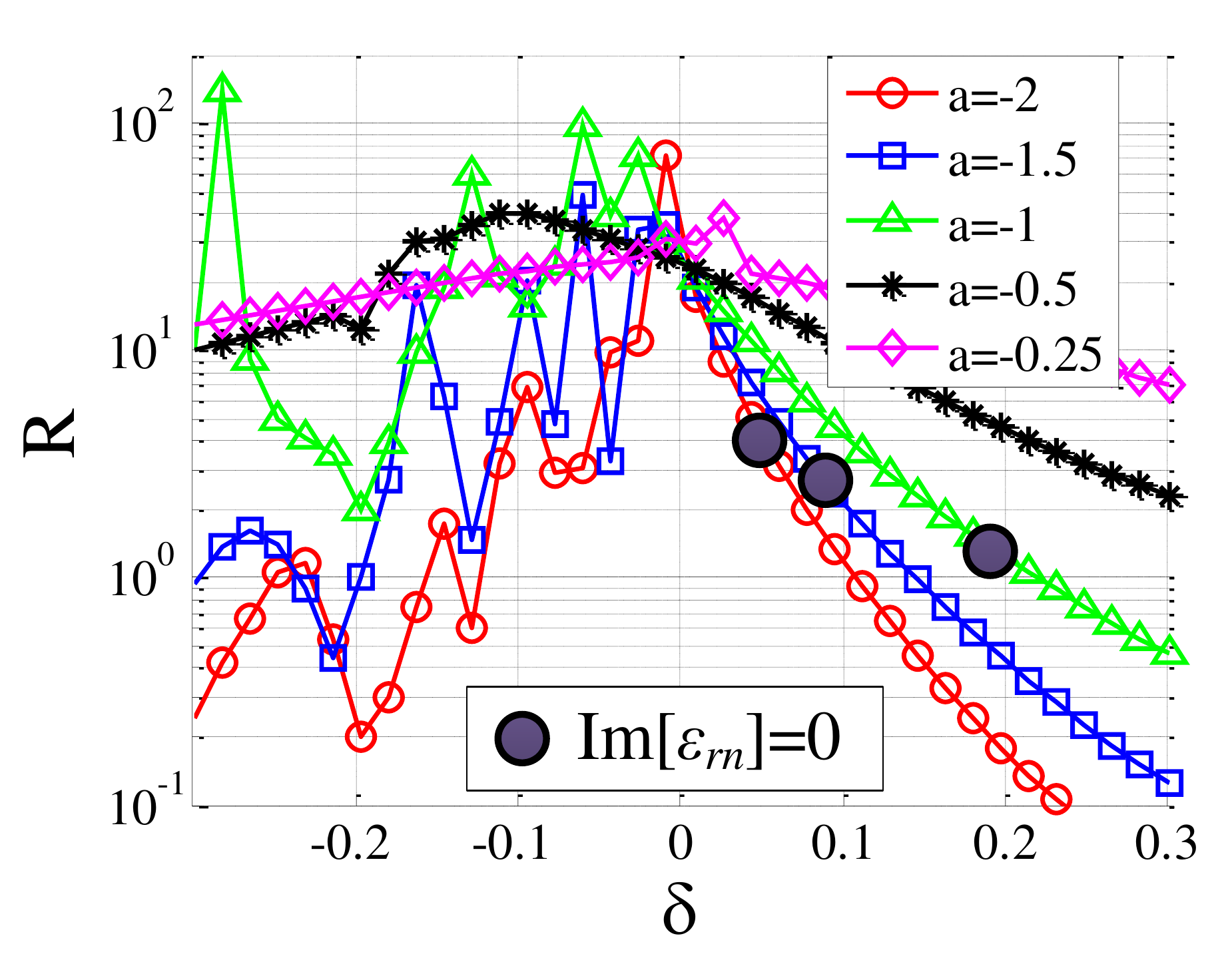}
\label{fig:Fig5a}}
\subfigure[]{\includegraphics[scale =0.43]{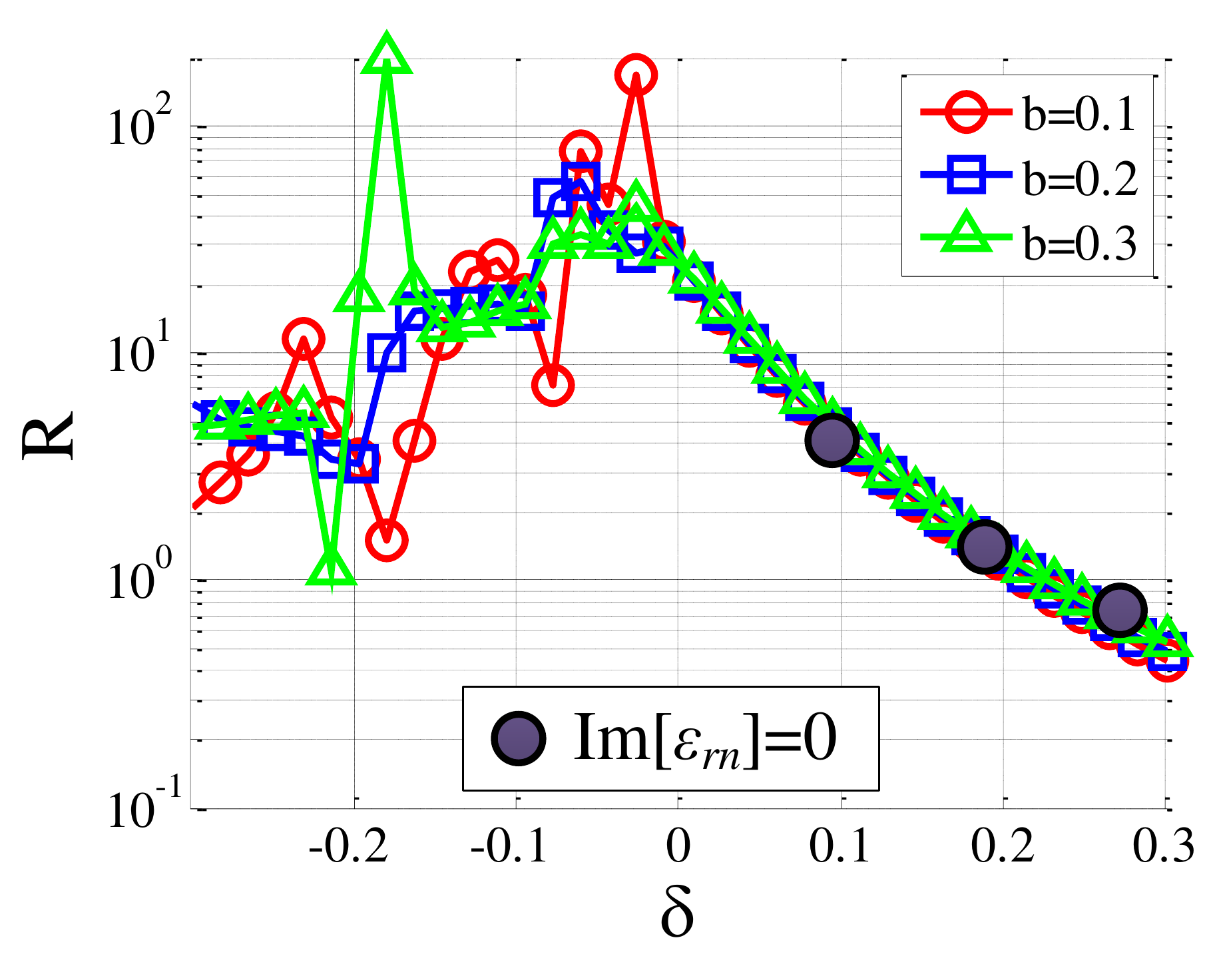}
\label{fig:Fig5b}}
\caption{The radiation enhancement ratio $R$ as function of the perturbation parameter $\delta=\Im[1/\e_{rt}-\e_{rn}]$ for: (a) various real part of transversal components $a=\Re[\e_{rt}]=\Re[\mu_{rt}]$ (with $b=0.2$) and (b) various imaginary parts $b=-\Im[\e_{rt}]=-\Im[\mu_{rt}]$ (with $a=-1$). Plot parameters: $r=\lambda_0/200$, $N=80$, $g=\lambda_0/20$, $L=3\lambda_0$, $\theta=90^{\circ}$.}
\label{fig:Figs5}
\end{figure}

In Fig.~\ref{fig:Fig5a}, we depict the variations of the radiation enhancement $R$ with respect to the perturbation parameter $\delta=\Im\left[\frac{1}{\e_{rt}}-\e_{rn}\right]=\Im\left[\frac{1}{\mu_{rt}}-\e_{rn}\right]$ for several values of the real parts $a$ of the transversal constituent components. When $\delta>0$, namely, when the structure is overall passive, the effect of the radiation vessels becomes weaker and weaker for increasing $\delta$, which is also obvious from Fig.~\ref{fig:Fig4a}. Note, however, that when the real part of the transverse permittivity approaches zero remaining negative, radiated power enhancement remains huge even for rather large positive $\delta$, that is, for rather high overall  losses in the system. On the other hand, when $\delta<0$, the structure is overall active and the whole slab acts as an additional power source; that is why the variations are sharper and more parameter-dependent. In particular, $R$ possesses substantially oscillating and, on the average, much higher values when $\delta<0$. The fluctuations are weaker and the output more stable when $a$ is negative but close to zero. Again one can observe the behavior of the system along the ``ultimate passivity limit'' (purple dots) which indicate once again that the functions $R=R(\delta>0)$ and $R=R(a<0)$ are decreasing.

In Fig.~\ref{fig:Fig5b} the change of $R=R(\delta)$ is shown for various $b=\Im[\e_{rt}]=\Im[\mu_{rt}]$. The radiation enhancement is almost independent from the imaginary part $b$ in the passive scenario, while, similarly to Fig.~\ref{fig:Fig5a}, shaky response is observed when $\delta<0$. It appears that when the system is overall active, one can find specific narrow intervals of $\delta$ where extremely high radiation is achieved regardless of the inherent losses $b$ along the transversal directions.

\begin{figure}[ht]
\centering
\subfigure[]{\includegraphics[scale =0.54]{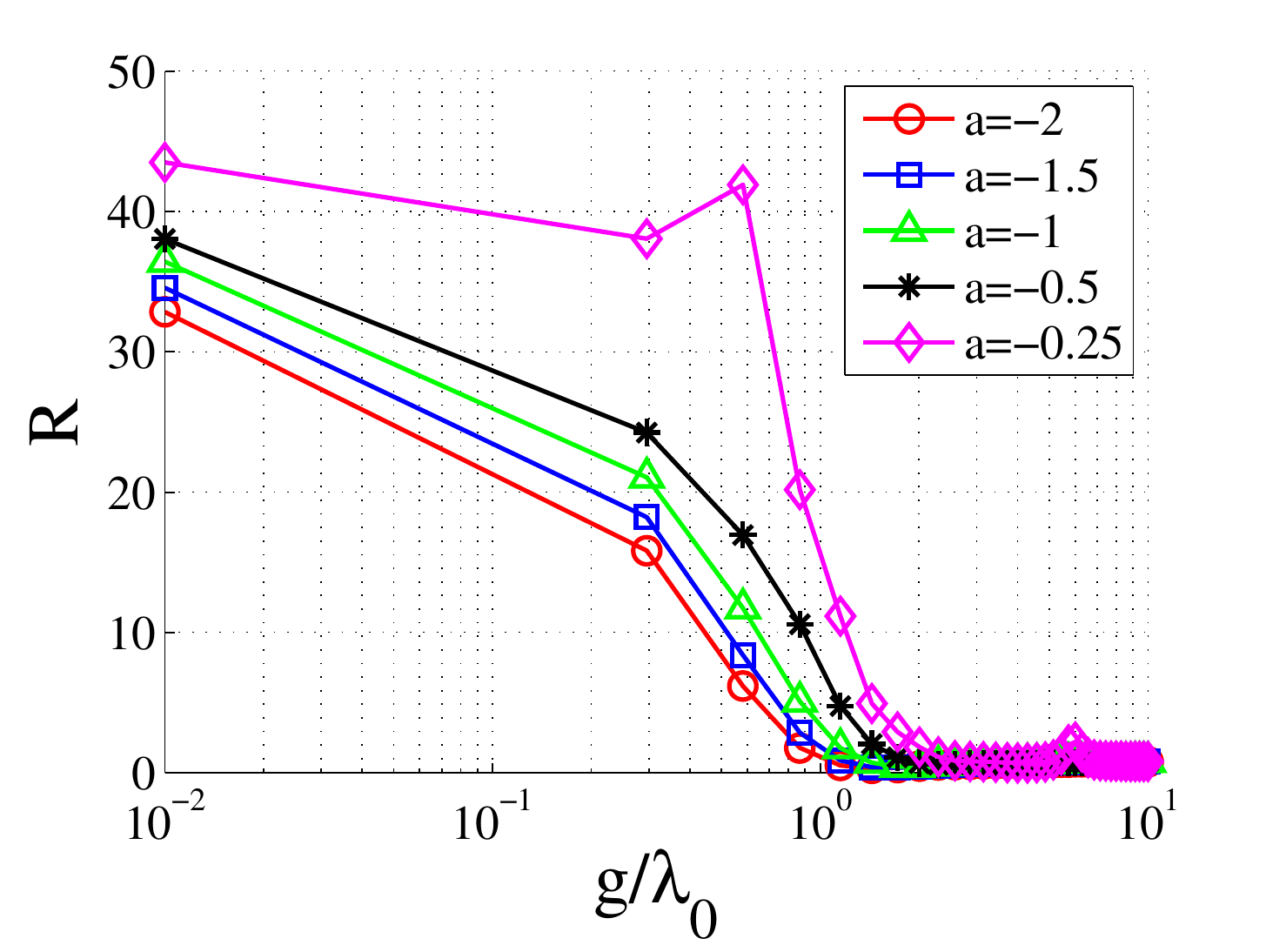}
\label{fig:Fig6b}}
\subfigure[]{\includegraphics[scale =0.54]{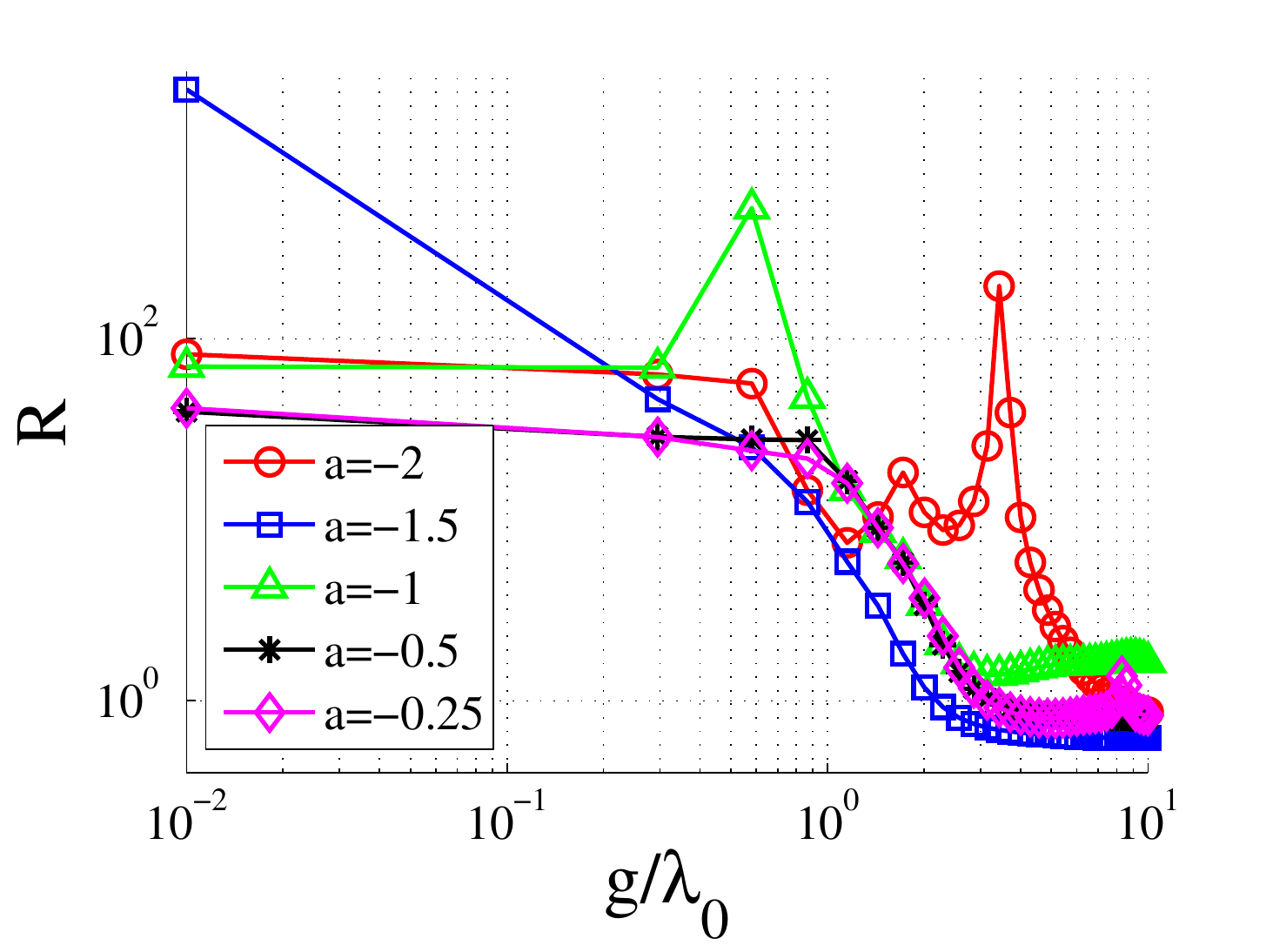}
\label{fig:Fig7b}}
\caption{The radiation enhancement ratio $R$ for various values of the real part of transversal components $a=\Re[\e_{rt}]=\Re[\mu_{rt}]$ as a function of the electrical distance of the source from the interface $g/\lambda_0$ ($L=3\lambda_0$). (a) Overall passive CML with $\delta=0.02$; (b) Overall active CML with $\delta=-0.02$.  Plot parameters: $b=0.2$,  $r=\lambda_0/200$, $N=80$, $\theta=90^{\circ}$.}
\label{fig:Figs6}
\end{figure}

In Figs.~\ref{fig:Figs6}, we identify the influence of the  location of the primary source in representative passive and active scenarios ($\delta=\pm 0.02$). In Fig.~\ref{fig:Fig6b} we can see that the radiation falls rapidly as the primary source gets distant from the air-CML slab interface, because the evanescent part of the exciting field gets weaker. However, especially in the active case shown in Fig.~\ref{fig:Fig7b}, radiation enhancement remains significant even when the distance to the source is much larger than the wavelength. As indicated above, in the active case the enhancement factor $R$ takes, on the average, higher values and exhibits a less monotonic behavior as a function of the geometrical and material parameters of the configurations.

\subsection{Radiation Patterns}
Apart from the macroscopic insight offered by the radiation-enhancement metric $R$, one can understand many features by observing the azimuthal variations of the far-field patterns for the introduced radiation-enhancing mirrors. The represented quantities are normalized by $|h_{\rm inc}(0)|=k_0\omega q/4$. The incident field in the far region $h_{\rm inc}(\f)$ is evaluated by (\ref{IncidentFarField}), the background field $h_{\rm inc}(\f)+h_{\rm sec}(\f)$ is given by (\ref{SecondaryFarField}), and the total field in the presence of the cylinders $h_{\rm inc}(\f)+h_{\rm sec}(\f)+h_{\rm scat}(\f)$ is computed using (\ref{ScatteredFarField}). 

\begin{figure}[ht]
\centering
\subfigure[]{\includegraphics[scale =0.54]{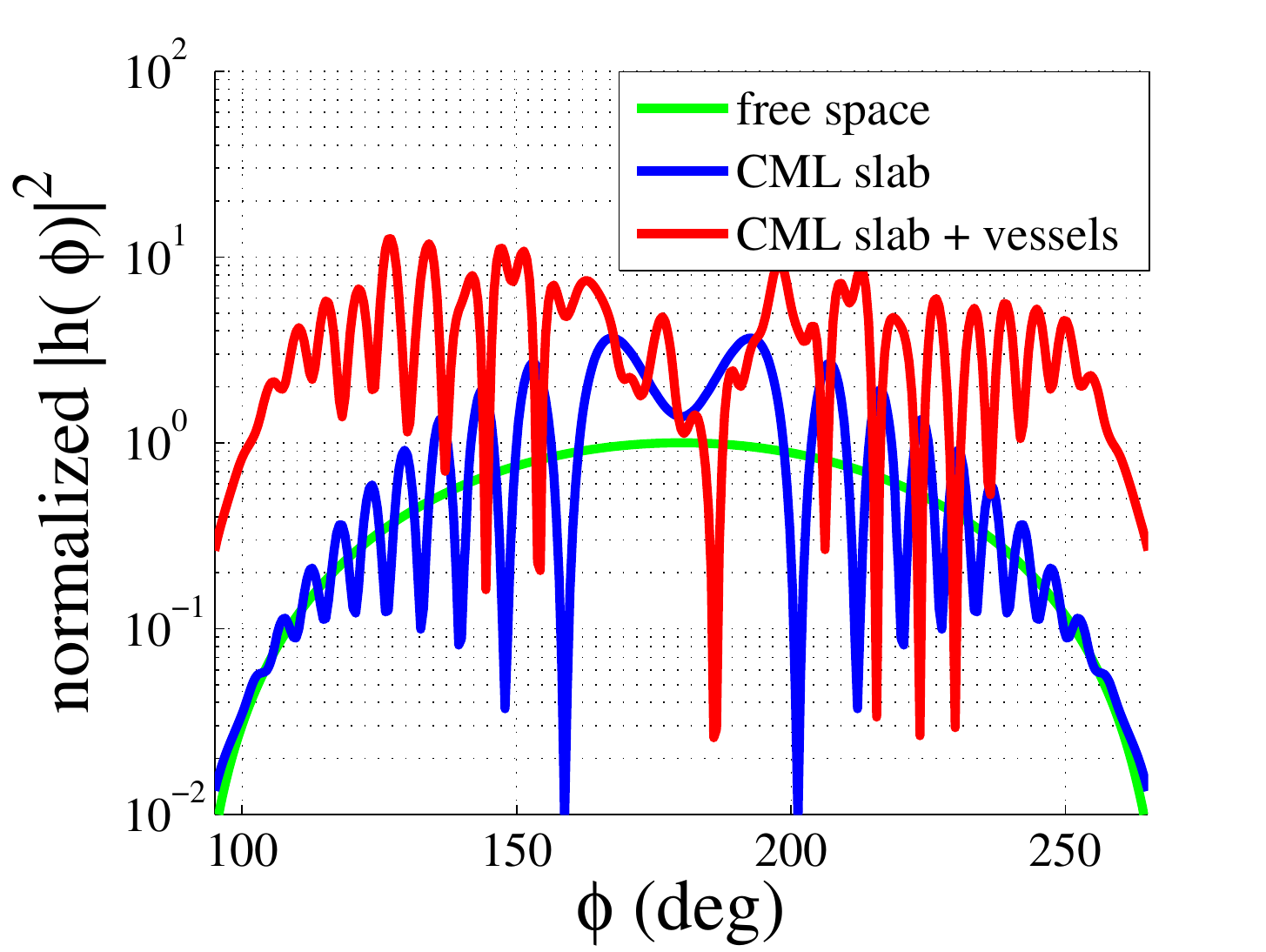}
\label{fig:Fig9a}}
\subfigure[]{\includegraphics[scale =0.54]{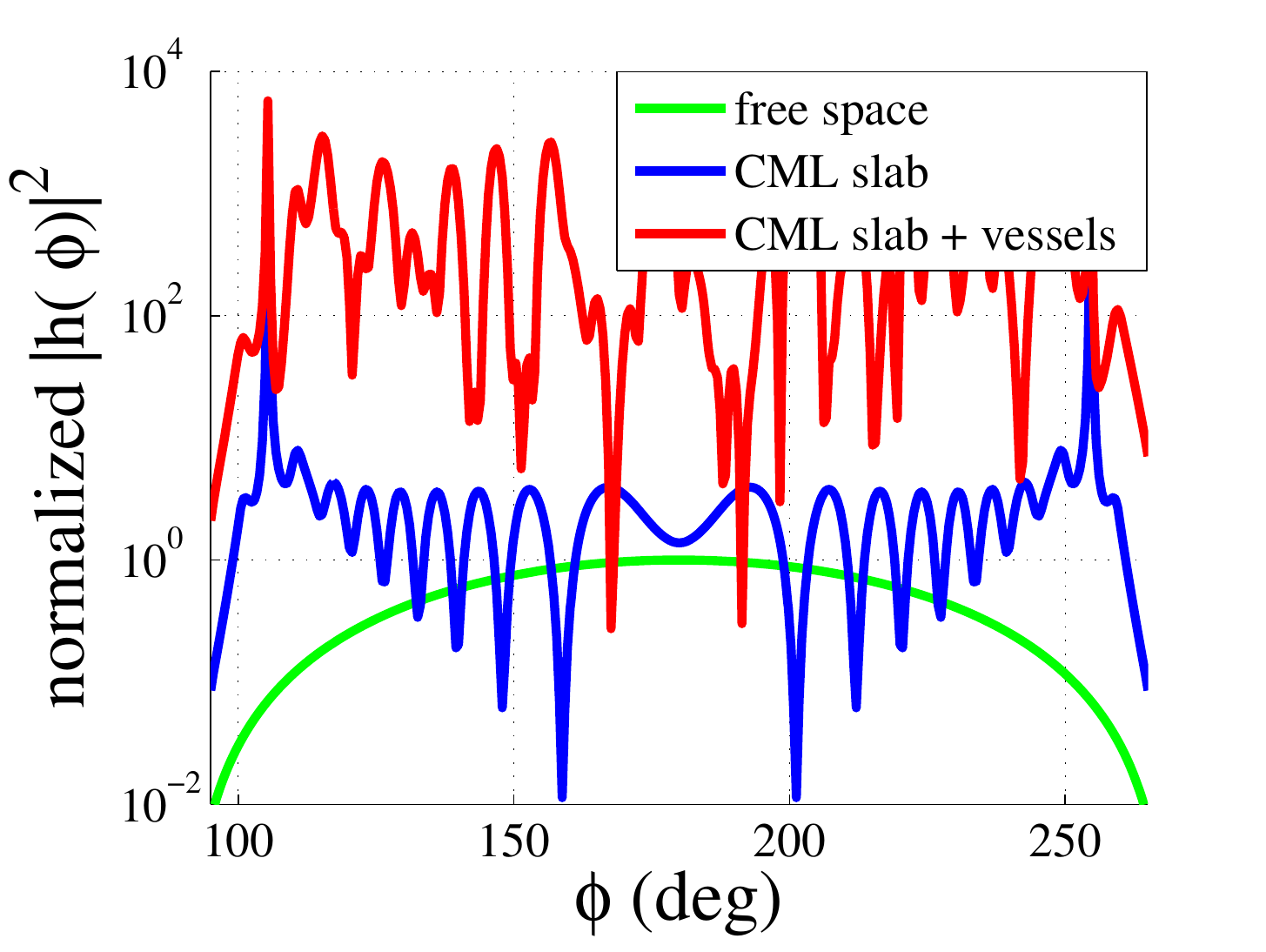}
\label{fig:Fig9b}}
\caption{The azimuthal profiles $|h(\f)|^2$ (normalized by $|h_{\rm inc}(0)|^2=k_0^2\omega^2q^2/16$) of the incident field, the background field and the total field as functions of angle $\f$ for: (a) a passive scenario ($\delta=+0.01$) and (b) an active scenario ($\delta=-0.01$). Plot parameters: $a=-2$, $b=0$, $L=3\lambda_0$, $g\lambda_0/10$, $r=\lambda_0/100$, $N=80$,  $\theta=90^{\circ}$.}
\label{fig:Figs9}
\end{figure}

In Figs.~\ref{fig:Figs9}, we illustrate two characteristic cases: one passive (Fig.~\ref{fig:Fig9a}) and one active (Fig.~\ref{fig:Fig9b}). The far-field patterns are represented in the case that the grounded CML slab is nearly fully reflecting propagating waves (we select the dissipation parameter $b=0$ and the CML loss factor $\delta$ is small). The three curves compare the far-field pattern of the primary source into free space (green), the pattern for the CML slab without perturbing pins (blue), and the CML slab with radiation enhancing vessels (red). The green curve is simply the pattern of a dipole line source, with the maximum in the broadside direction (in both Figs.~\ref{fig:Figs9}). CML slab without pins basically acts as a reflector for the propagating part of the incident spectrum and its response gets substantially enhanced along the grazing-angle directions ($\f\cong 90^{\circ}, 270^{\circ}$) for the active scenario. We can see that for the passive mirror the maximum enhancement of the field strength equals 2 (four-fold in terms of power), which takes place for directions along which the reflected field sums up in phase with the field of the primary source. This value is the maximal possible value of the reflected field from any ideally reflecting planar mirror (with arbitrary reflection phase). We clearly see that the radiation vessels provide an additional radiation channel via the evanescent part of the spectrum, and the radiated power is strongly enhanced, well above the fundamental limit for any lossless mirror.

\begin{figure}[ht]
\centering
\subfigure[]{\includegraphics[scale =0.54]{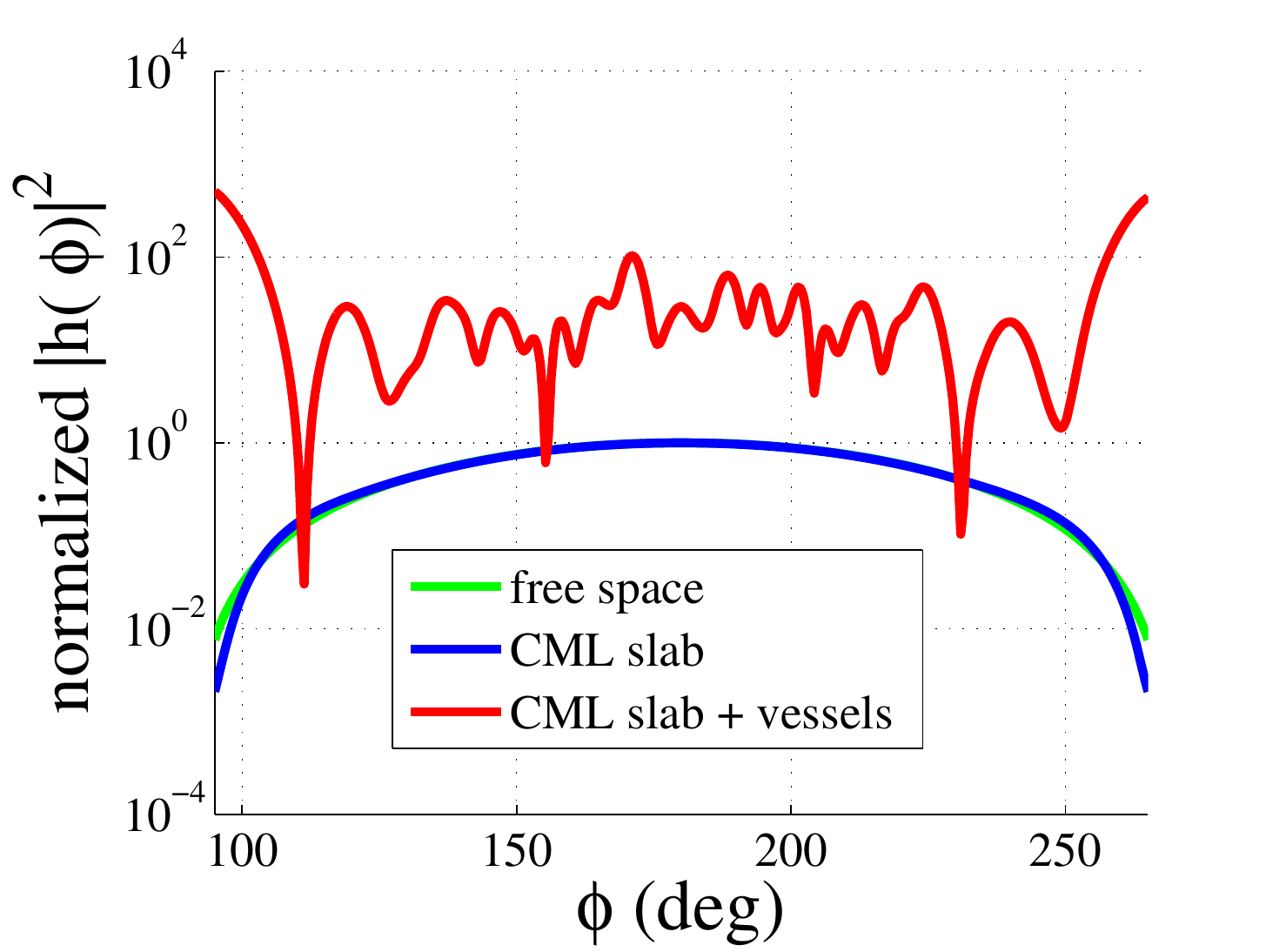}
\label{fig:Fig8a}}
\subfigure[]{\includegraphics[scale =0.54]{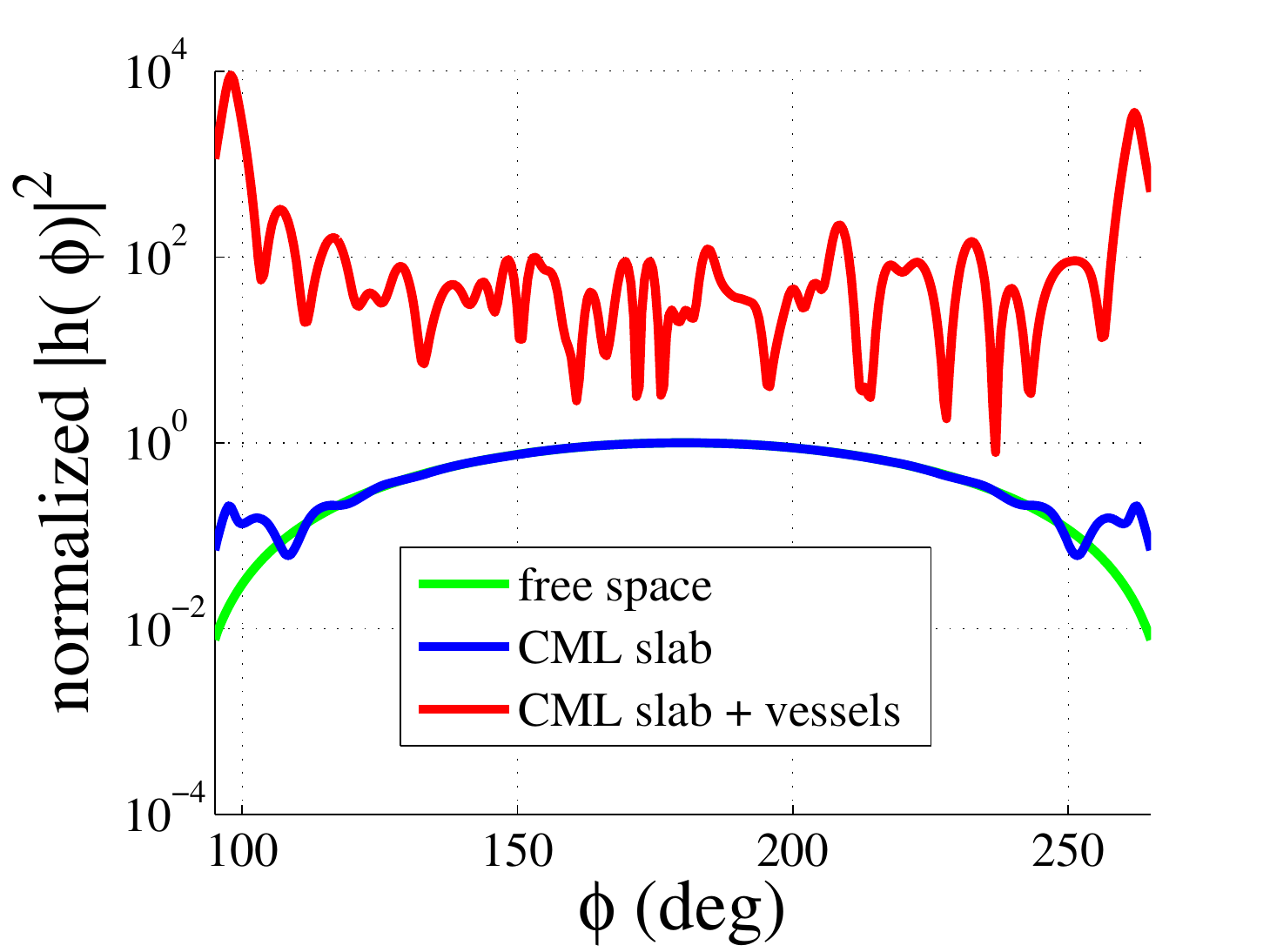}
\label{fig:Fig8b}}
\caption{The azimuthal profiles $|h(\f)|^2$ (normalized by $|h_{\rm inc}(0)|^2=k_0^2\omega^2q^2/16$) of the incident field, the background field and the total field for: (a) $a=-0.25$, $\delta=0.02$, $L=2.557\lambda_0$, $g=\lambda_0/20$ ($R\cong 84$) and (b) $a=-1$, $\delta=-0.02$, $L=3\lambda_0$, $g=0.0665\lambda_0$ ($R\cong 525$). Plot parameters: $b=0.2$, $r=\lambda_0/200$, $N=80$,  $\theta=90^{\circ}$.}
\label{fig:Figs8}
\end{figure}

In Figs.~\ref{fig:Figs8}, we represent the results also for a passive (Fig.~\ref{fig:Fig8a}) and an active (Fig.~\ref{fig:Fig8b}) case which correspond to high radiation enhancement $R$. Losses in the CML are present ($b=0.2$), so that the reflections of the propagating part of incident waves are weak. That is why we see that the blue and green curves nearly coincide for the passive CML. The chosen value of $\delta=0.02$ leads to an enhancement in radiation by a factor of $R\cong 84$. In this case, it is apparent that the far-field response of the CML slab without the radiation enhancing cluster is almost identical to the incident field, which is anticipated from (\ref{SecondaryFarField}). However, when one puts the randomly distributed vessels in the near field, the output power of the antenna gets significantly amplified (and the pattern becomes asymmetric with respect to $\f=180^{\circ}$). In Fig.~\ref{fig:Fig8b}, corresponding to an active CML, we use $\delta=-0.02$ and the enhancement is huge in all directions (the overall radiated power enhancement factor $R\cong 525$). 

\section{Conclusions}
It is well known that infinite homogeneous planar surfaces can fully reflect electromagnetic waves in the limit of negligible losses. In this case, the amplitude of waves radiated by a source near the mirror can be doubled as compared with the incident waves. This surface is thought to be ``ideally shiny." On the other hand, the reflection coefficient from planar surfaces can be in principle made zero for all incident propagating waves (any polarization and any incident angle). In this case, all power of incident propagating plane waves is absorbed and the surface is ``ideally black," absorbing maximum power and, reciprocally, emitting maximal heat power according to the Plank law. In both these scenarios, evanescent waves do not participate in power exchange between far-zone external sources and the material body. 

In this paper, we have shown that perturbing the surface of an infinite planar surface which maintain resonant surface modes we can in principle realize a planar reflector which reflects more power than any ideally reflecting planar surface. Due to perturbations, surface modes couple to propagating plane waves and create additional channels for power exchange via evanescent fields. Such perturbed resonant surface extracts extra power from a near sources and sends that into space. The amplitude of the reflected field can be orders of magnitude larger than the maximal value of 2 for any usual lossless mirror. Making the perturbation lossy, it can become possible to overcome the black-body limit even for planar surfaces. In this scenario, the black body absorbs nearly all propagating waves, while the perturbations provide coupling between the resonant surface modes and an additional energy sink via evanescent modes.     

The perturbations are random at the wavelength scale and non-resonant. In this configuration, the surface-averaged currents induced in the perturbation objects (which could partially reflect propagating waves and compromise their absorption in the CML body) are small because they couple to non-resonant propagating modes of the absorbing or reflecting body. On the other hand, current components which vary fast on the wavelength scale can be huge because they couple to highly  resonant CML body. These spatially inhomogeneous resonant currents on the perturbations provide additional channels for power exchange between the body and free-space wave modes, allowing stronger reflections than from an ideal reflector or more absorption than in the ideal black body.      

Analyzing the far-zone radiation patterns we see that there is an analogy between the revealed phenomena and superdirectivity of antennas \cite{Sch}. Superdirective radiators can create a narrow beam with the directivity higher than that of the same configuration which is uniformely excited \cite{Hansen}. However, in the configuration which we have introduced here, it appears that the planar reflector sends superdirective beams nearly everywhere (pattern oscillations are determined by random positions of radiation vessels). There is also a connection of the revealed phenomena to the concept of perfect lens as a slab of a lossless double-negative material \cite{Pendry}. The perfect lens operation also exploits resonance of surface modes at an interface between free space and a double-negative material. In the perfect lens concept, high-amplitude reactive fields at the entry interface are focused behind the lens thanks to interactions between resonant modes of the two parallel surfaces of the lens. In papers \cite{SphericalPaper,MAXABS} it was shown how the reactive energy of the resonant surface modes can be fully absorbed. Here we have shown that this energy can be launched into space, creating super-reflectors and far-field superemitters.

Although in this paper we have considered a particular realization of surface perturbations in form of a random array of thin cylinders, the concept is general and the surface can be perturbed in many various ways, for instance simply making the surface rough at the appropriate wavelength scale. Likewise, the surface does not have to be planar nor infinite: properly perturbing the surface of a finite-size conjugately matched body we can dramatically enhance its coupling to electromagnetic fields in space. Discussed  super-emission and super-absorption phenomena can potentially enable new approaches to optimizing wireless transfer of energy or information and in radiative heat transfer management.

\end{document}